\documentclass[fleqn,usenatbib,useAMS]{mnras}

\usepackage{graphicx}	% Including figure files
\usepackage{amsmath}	% Advanced maths commands
\usepackage{amssymb}	% Extra maths symbols
\usepackage{multicol}        % Multi-column entries in tables
\usepackage{bm}		% Bold maths symbols, including upright Greek
\usepackage{pdflscape}	% Landscape pages

\usepackage[T1]{fontenc}
\usepackage{ae,aecompl}

\usepackage{newtxtext,newtxmath}

\title[Two substellar survivor candidates]{Two substellar survivor candidates; one found and one missing}

\author[N. Walters et al.]{N. Walters,$^1$\thanks{E-mail: nikolay.walters.15@ucl.ac.uk}
J. Farihi,$^{1}$
T. R. Marsh,$^{2}$\thanks{Deceased.}
E. Breedt,$^{3}$
P. W. Cauley,$^{4}$
T. von Hippel,$^{5}$
J. J. Hermes$^{6}$
\medskip
\\
$^1$Department of Physics and Astronomy, University College London, London WC1E 6BT, UK\\
$^2$Department of Physics, University of Warwick, Coventry CV4 7AL, UK\\
$^3$Institute of Astronomy, University of Cambridge, Madingley Road, Cambridge CB3 0HA, UK\\
$^4$Laboratory of Atmospheric and Space Physics, University of Colorado Boulder, Boulder, CO 80309, USA\\
$^5$Department of Physical Sciences, Embry-Riddle Aeronautical University, Daytona Beach, FL 32114, USA\\
$^6$Department of Astronomy, Boston University, 725 Commonwealth Ave., Boston, MA 02215, USA
}
\date{Accepted XXX. Received YYY; in original form ZZZ}
\begin{document}

%\label{firstpage}

\maketitle

\begin{abstract}

This study presents observations of two possible substellar survivors of post-main sequence engulfment, currently orbiting white dwarf stars.  Infrared and optical spectroscopy of GD\,1400 reveal a 9.98\,h orbital period, where the benchmark brown dwarf has $M_2=68\pm8$\,M$_{\rm Jup}$, $T_{\rm eff}\approx2100$\,K, and a cooling age under 1\,Gyr.  A substellar mass in the lower range of allowed values is favoured by the gravitational redshift of the primary.  Synthetic brown dwarf spectra are able to reproduce the observed CO bands, but lines below the bandhead are notably overpredicted. The known infrared excess towards PG\,0010+281 is consistent with a substellar companion, yet no radial velocity or photometric variability is found despite extensive searches.  Three independent stellar mass determinations all suggest enhanced mass loss associated with binary evolution, where the youngest total age for an isolated star is $7.5\pm2.5$\,Gyr.  A possible solution to this conundrum is the cannibalization of one or more giant planets, which enhanced mass loss post-main sequence, but were ultimately destroyed.  PG\,0010+281 is likely orbited by a debris disk that is comfortably exterior to the Roche limit, adding to the growing number of non-canonical disks orbiting white dwarfs.  At present, only L-type (brown) dwarfs are known to survive direct engulfment during the post-main sequence, whereas T- and Y-type substellar companions persist at wide separations.  These demographics indicate that roughly 50\,M$_{\rm Jup}$ is required to robustly avoid post-main sequence annihilation, suggesting all closely-orbiting giant planets are consumed, which may contribute to mass loss and magnetic field generation in white dwarfs and their immediate progenitors.

\end{abstract}


\section{Introduction}

It is now well established that white dwarfs often host planetary systems (e.g. \citealt{2016Veras}), largely demonstrated by metal pollution and circumstellar matter \citep{2016Farihi}, and more recently transiting debris clouds as well as a few giant planet candidates \citep{2015Vanderburg, 2019Gansicke, 2020Vanderburg, 2022Farihi}.  And while sun-like (FGK) stars are well characterized over a wide range of planetary masses and orbital radii \citep{2017Millholland, 2019Hsu, 2020Kunimoto, 2021Jin}, main-sequence stars of intermediate mass (A-type and similar) are generally not amenable to conventional exoplanet detection techniques such as precision radial velocities and transits.  However, their white dwarf descendants, which can be formed within 1\,Gyr for stars above 2\,M$_\odot$ \citep{2012Mowlavi}, are -- at least in some ways -- easier substellar companion search targets.  For example, in addition to the powerful sensitivity of atmospheric pollution \citep{2007Zuckerman}, brown dwarf and high-mass planetary companions can be detected through infrared excess \citep{2019Rebassa}, and transit depths can be enhanced by a factor of over 100.  Moreover, statistically speaking, extant white dwarf planetary system hosts likely outnumber their progenitors systems; there are 20 white dwarfs within 10\,pc, and only 12 stars earlier than G-type \citep{2021Reyle}.  Complicating the picture somewhat, is the evolution of planetary systems from the main sequence through the two giant branches, and into the white dwarf phase.  What survives, and are the substellar survivors altered?

Low-mass stellar and substellar companions to white dwarfs provide important insight into the above question, albeit at the higher mass end of possible survivors.  Because brown dwarfs are not expected to be altered during common envelope evolution, low-mass ratio binaries carry information on the initial (and companion) mass functions, including an avenue for exploration of the known deficit of brown dwarf companions to main-sequence stars \citep{2004McCarthy,2006Grether}. Furthermore, the study of bound companions provides more information than compared to isolated objects, as a binary is the smallest possible `cluster'.  For systems with a white dwarf primary, there will be a clear age constraint from its cooling age, which can be applied directly to the secondary, and where state-of-the-art Bayesian methods can provide a decent constraint on the total system age \citep{2006vonHippel, 2022Moss}.

The frequency of white dwarfs hosting a brown dwarf companion is also low, where an extensive near-infrared imaging survey measured the fraction to be below 0.5 per cent \citep{2005Farihi_a}.  This estimate agrees with later, near-infrared photometric surveys covering a large fraction of the sky \citep{2011Steele, 2011Girven}, and a subsequent \textit{WISE} study \citep{2011Debes}.  Although possibly incomplete in terms of its low-mass brown dwarf companion census, the 40\,pc sample of white dwarfs contains only two substellar companions; the prototype L dwarf GD\,165B \citep{1988Becklin}, and the first sub-T dwarf candidate WD\,0866-661B \citep{2011Luhman}\footnote{These astrophysical firsts were detected as companions to white dwarfs, underscoring their utility and impact.  Combined with the prototype T dwarf Gl\,229B, discovered as a companion to an M dwarf \citep{1995Oppenheimer}, low-luminosity stars have been paramount in the discovery and establishment of substellar spectral classes.}.  This yields a fraction of brown dwarf companions close to 0.2 per cent \citep{2019Gentile}.

From both theory and observation, low-mass stellar and substellar companions to white dwarfs form two distinct populations, differentiated by orbital separation and separated by a period gap \citep{2006Farihi, 2011Nebot}. The existence of the two populations is consistent with the expectations of stellar evolution \citep{2010Nordhaus}, where close pairs have experienced a phase of common envelope evolution \citep{1976Paczynski} that has dramatically decreased their initial orbits at formation.  In wide binaries, the secondary has migrated outwards by a factor directly proportional to the mass lost during the red giant (RGB) and asymptotic giant (AGB) branches of the host \citep{1924Jeans}.

In the case where a substellar companion persists, after the engulfment by its first ascent giant or asymptotic giant host star, the brown dwarf is expected to be strongly irradiated during the first few Gyr of white dwarf cooling.  Such systems can provide empirical data that can be compared to existing models of irradiated hot Jupiter atmospheres \citep{2022Lee, 2020Tan}.  The lowest-mass survivors of the common envelope provide direct testimony of planetary survival under these conditions and thus provide useful benchmarks.  An example might be the recently reported giant planet candidate transiting a white dwarf \citep{2020Vanderburg}, where the expected effects of a common envelope (e.g.\ atmospheric ablation, orbital dissipation) may imply the transiting object migrated much later to its current orbital position, or was originally more massive.  Despite almost a thousand candidate post-common envelope systems \citep{2021Kruckow}, this phase of binary evolution remains a major modelling challenge \citep{2011Ivanova}. 

This paper presents new observations of GD\,1400 and PG\,0010+281, where both are suspected of being white dwarf - brown dwarf binaries that underwent post-common envelope evolution.  Orbital and binary parameters are successfully obtained for both components of GD\,1400, whereas only companion limits are obtained for PG\,0010+281.  In Section~\ref{sec:obs} new and archival observations are described, followed by two sections of analysis and results focussing on the separate science targets: Section~\ref{sec:data1} for GD\,1400, and Section~\ref{sec:data2} for PG\,0010+281, including all findings, followed by concluding remarks.

\section{Observations and Data}\label{sec:obs}

This section is divided over the two science targets, and describes new and archival observations of GD\,1400 and PG\,0010+281, where a summary is provided in Table~\ref{tab:obs_sum}. 

\subsection{GD\,1400}

GD\,1400 was observed with five instruments to constrain its binary orbit and companion mass.  The order presented below follows the chronology of these various datasets.

\subsubsection{High-resolution imaging}

Subsequent to the spectroscopic discovery of GD\,1400B in 2004 -- the second known substellar companion to any white dwarf \citep{2004Farihi} -- two high-resolution imaging observations were undertaken in an attempt to spatially resolve the brown dwarf from the white dwarf primary. \textit{Hubble Space Telescope} (\textit{HST}) observations of GD\,1400 were executed on 2005 June 5 with the Advanced Camera for Surveys (ACS; \citealt{1998Ford}) as part of Cycle 13 Snapshot programme 10255 \citep{2006Farihi}. A four-position imaging sequence was executed using the High Resolution Camera and the F814W filter (similar to Cousins $I$ band) for a total exposure time of 1080\,s. The resulting fully-processed and multidrizzled science image revealed no evidence for the companion (including stellar image elongation), although point spread function subtraction was not performed. While these space-based data are sensitive to sub-arcsecond spatial separations, the expected $\Delta I\sim 6$\,mag difference between the white dwarf and its L dwarf companion \citep{2002Dahn,2005Farihi_b} allows for the possibility that such a binary could remain unresolved at the 0.1\,arcsec level.

\begin{figure}
\includegraphics[width=\columnwidth]{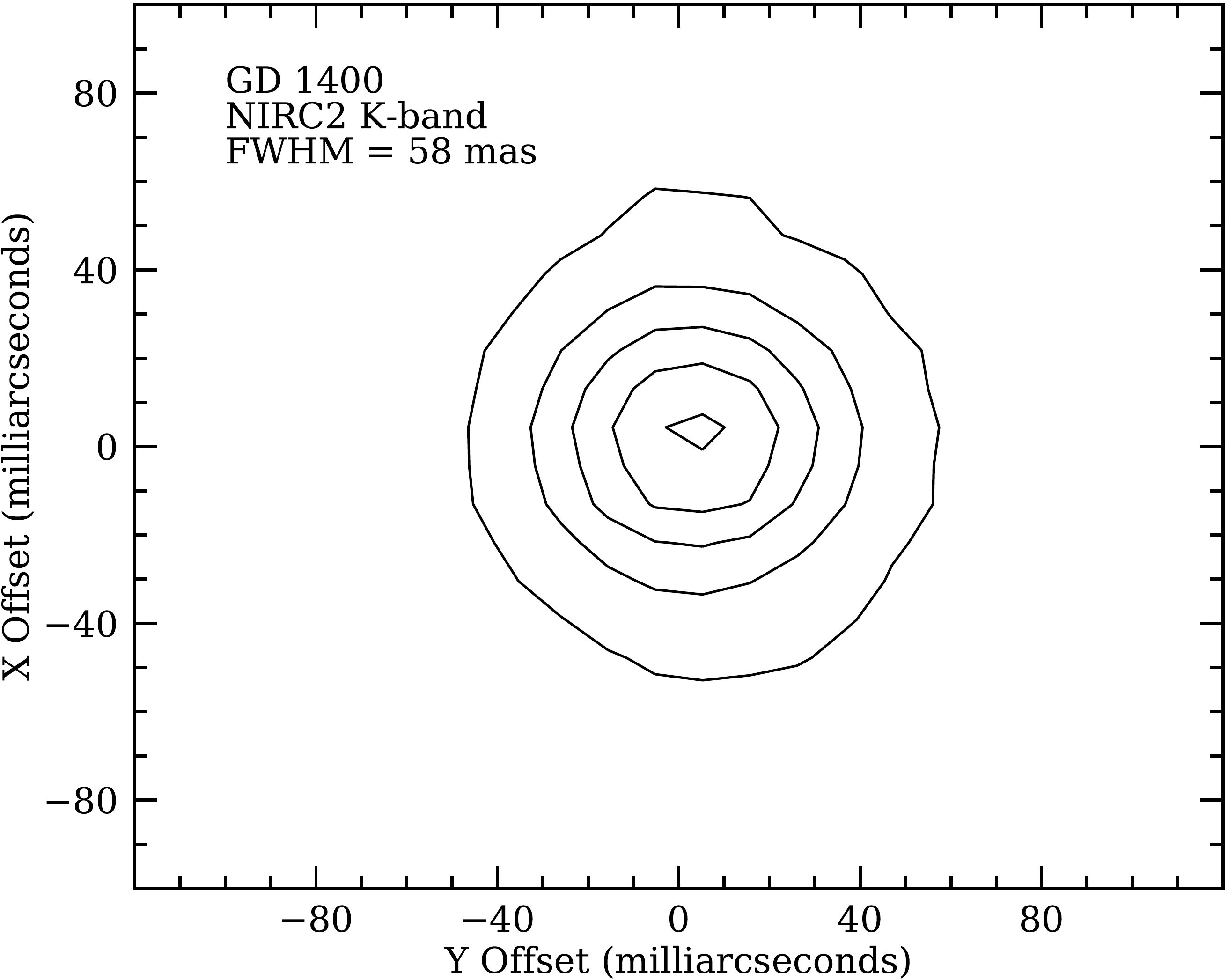}
\vskip 0mm
\caption{Flux density contours for the Keck NIRC2, $K$-band adaptive optics image with the highest Strehl ratio. The five evenly-spaced contours are plotted from 1000 to 5000 counts, and the FWHM is measured to be 58\,mas.  For a pair of equally luminous binary components, a separation of approximately 20\,mas ($\approx0.9$\,au) can be ruled out.}
\label{fig:gd1400_ao}
\end{figure}

Laser guide star adaptive optics observations of GD\,1400 were obtained on 2005 September 10 on Mauna Kea using the Keck II telescope and the Near-Infrared Camera 2 (NIRC2; \citealt{2006vanDam,2006Wizinowich}) as part of H26aN2 program. Four images of GD\,1400 were taken in the $K$ band and narrow field camera (0.01\,arcsec$^2$\,pixels), each utilising 20\,s exposures with 2 co-adds. Because the white dwarf and brown dwarf should have roughly equal flux density at this wavelength \citep{2004Farihi}, the image with the best corrections and highest Strehl ratio was examined for any evidence of duplicity. Contours for this signal-to-noise (S/N) $>300$ image are shown in Figure~\ref{fig:gd1400_ao}. The point spread function is sufficiently symmetric that an equally luminous binary separated by $1/3$ the full width at half maximum (FWHM) can be ruled out, implying that the pair must have a projected separation within 0.02\,arcsec.  The latest \textit{Gaia} DR3 observations \citep{GaiaDR3} place the system at 46.3\,pc, thus limiting the projected separation on the sky to less than 0.9\,AU.  Armed with these data that suggest a likely orbital period of less than around 1\,yr, a programme to monitor the radial velocity of GD\,1400A was carried out in 2006.

\subsubsection{Optical spectroscopy of GD\,1400A}

High-resolution optical spectroscopy of the white dwarf was performed using the Ultraviolet and Visual Echelle Spectrograph (UVES; \citealt{2000Dekker}) mounted at the European Southern Observatory (ESO) Very Large Telescope (VLT). A total of 34 spectra were obtained covering H$\upalpha$ and the higher Balmer lines; two of these spectra were acquired under the ESO SPY survey \citep{2003Napiwotzki} several years prior to the dedicated follow up observations.  The remaining 32 spectra were observed as part of programme 077.D-0673 between 2006 July 06 and September 03, using the same UVES setup as the SPY survey: a 2.1\,arcsec slit and $2\times2$ binning on the array. The setup provided spectral resolving power of $R\approx18\,500$ \citep{2001Napiwotzki}.

The 2006 observations were performed in service mode, in seeing no worse than 1.4\,arcsec, for a total exposure time of 1200\,s per visit. Each observing visit was split into $2\times600$\,s exposures to avoid orbital smearing in case the period was shorter than a few hours.  On 2006 July 18, no data in the blue arm were taken, resulting in four missing exposures covering 3250 to 4500\,\AA, which is attributed to human or instrumental error (no reasons are provided in the night log).  All data were extracted in a semi-automatic way using the ESO Reflex Environment ({\sc esoreflex}; \citealt{2013Freudling}) and UVES pipeline version 6.1.6 \citep{2000Ballester}. The standard recipes were used to optimally extract and wavelength calibrate each spectrum.  The resulting average S/N was estimated to be 31 in the H$\upbeta$ region and 29 in the H$\upalpha$.

%%%TABLE 1%%%
\begin{table}
\centering
\caption{Journal of new and archival observations with approximate dates.}
\label{tab:obs_sum}
\begin{tabular}{ccc}

\hline
Telescope \&    &Data Type                  &Dates of\\
Instrument      &                           &Observation\\

\hline

\multicolumn{3}{l}{\bf GD\,1400:}
\smallskip\\

{\em HST} ACS   &F814W imaging              &2005 Jun 5\\
Keck NIRC2      &$K$-band AO imaging        &2005 Sep 10\\
VLT UVES        &Optical spectroscopy       &2000, 2006 Jul -- Sep\\
VLT ISAAC       &$K$-band spectroscopy      &2010 Oct\\
{\em TESS}      &Optical photometry         &2018, 2020 Sep -- Oct\\
\\

\multicolumn{3}{l}{\bf PG\,0010+281:}
\smallskip\\

Keck HIRES      &Optical spectroscopy       &2013 Sep\\
INT IDS         &"                          &2015 Jul\\
MMT Blue Chan   &"                          &2016 Jan 15\\
VLT UVES        &"                          &2017 Oct -- Nov\\
WHT ISIS        &"                          &2018 Aug -- Sep\\
Gemini GRACES   &"                          &2019 Nov\\
Keck NIRSPEC    &$K$-band spectroscopy      &2015 May 26\\
IRTF SpeX       &Near-infrared spectroscopy &2019 Oct 20\\
{\em TESS}      &Optical photometry         &2019 Oct -- Nov\\
ZTF             &"                          &2018 May -- 2021 Jan\\
\textit{WISE}   &Infrared photometry        &2010 -- 2020\\

\hline

\end{tabular}
\end{table}

\subsubsection{Infrared spectroscopy of GD\,1400B}

To measure the radial velocity of GD\,1400B, $K$-band time-series spectroscopy was carried out using the Infrared Spectrometer and Array Camera (ISAAC; \citealt{1999Moorwood}), a low- to medium-resolution, infrared spectrograph and imager previously mounted on the VLT.  The spectra were obtained over two consecutive nights under proposal 086.D-0030 on 2010 October 29 and 30 on a Rockwell 1024$\times$1024\,pixel$^2$ Hawaii array, using the short wavelength arm in spectroscopic mode with the medium-resolution grating. The slit width was set to the narrowest available option of 0.3\,arcsec, resulting in $R\approx8900$ over the spectral range $1.82-2.50\,\upmu$m.  Based on the discovery spectrum of the brown dwarf \citep{2004Farihi}, this setup was chosen to resolve individual transitions within the CO bands, and provide robust cross-correlation across velocity-shifted spectra.  A total of 62 exposures were obtained over the two nights, using 600\,s integrations and following the standard infrared ABBA pattern on the slit, i.e.\ nodding the object along the slit between two positions.

Spectral reduction was performed in \textsc{iraf} \citep{1986Tody,1993Tody}.  A bad pixel mask was constructed by identifying hot pixels in dark frames and cold pixels in flat field frames; this mask was then used to interpolate across all affected pixels in all science and calibration frames.  The two-dimensional data were flat fielded, then A$-$B and B$-$A pairs were created to remove the bright infrared sky, and for subsequent spectral extraction.  An A0V standard star (HD\,9132) was observed in a similar manner as the science target, and its strong spectral trace was used to extract the spectra of GD\,1400B.  For wavelength calibration, XeAr arcs were acquired multiple times throughout both nights to correct for any flexure in the optics, and to provide the best wavelength solution as a function of time.  Telluric correction and flux calibration were applied using the {\sc xtellcor\_general} routine within {\sc spextool} \citep{2003Vacca,2004Cushing}. The resulting spectra had an average S/N of 15.

%%%FIGURE 1%%%

\subsubsection{Time-series photometry}

GD\,1400 was observed by the {\em Transiting Exoplanet Survey Satellite} ({\em TESS}; \citealt{2015Ricker}), in Sector 3 from 2018 September 21 until 2018 October 17, and again in Sector 30 from 2020 September 23 until 2020 October 20.  The star was observed with camera 2, and on CCD chip number 3 in both sectors, under designation TIC\,164772507 (\textit{TESS} Mag = 15.2\,mag; \citealt{2018Stassun}).  \textit{TESS} utilises a red optical bandpass covering roughly $6000-10\,000$\,\AA, and each CCD has a $4096\times4096$\,pixel$^2$ on-sky area, which are read out at 2\,s intervals. The on-board computer then produces $11\times11$\,pixel$^2$ postage stamps centred at the target, with flux averaged over 120\,s. In some cases (for approximately 1000 targets per sector), data that are flux averaged over 20\,s are also available, as was the case for Sector 30 observations of GD\,1400.  

The availability of this higher cadence data is due to the white dwarf being specifically targeted as a known ZZ\,Ceti pulsator \citep{2003Fontaine}.  This is a stellar evolutionary phase common to all white dwarfs with hydrogen atmospheres, where an atmospheric instability results in non-radial pulsations (see \citealt{2022Romero} and references therein).  Given the approximate effective temperature and mass from prior publications, the star should be in the middle of the ZZ\,Ceti instability strip \citep{2015Tremblay}, where some stars have pulsation periods shorter than the standard 120\,s cadence \citep{2006Mukadam}.

Data products from both sectors were accessed from the \textit{TESS} Mikulski Archive for Space Telescopes. Light curves were extracted from the target pixel files by the \textit{TESS} Science Processing Operations Center pipeline \citep{2016Jenkins}. Pre-search Data Conditioning Simple Aperture Photometry (PDCSAP) light curves were used, since they have systematic errors removed, including error sources from the telescope and the spacecraft using co-trending basis vectors \citep{2012Smith,2012Stumpe}.  It should be noted that the pipeline corrects the flux of each target to account for crowding from other sources, where the flux of GD\,1400 is modelled as 0.97 of the total on average, so the corrections are minor. Pipeline defined apertures were used throughout for reliability and reproducibility provided by the PDCSAP. Across the two sectors 38\,259 120\,s measurements are available and 29\,552 of these have no anomalous flags.

\subsection{PG\,0010+281}

Three independent approaches were considered to identify or rule out possible substellar companions to PG\,0010+281 as suggested by its infrared excess \citep{2015Xu}.

\subsubsection{Optical spectroscopy}

Optical spectra were acquired (or extracted from archives) from several facilities as summarized in Table~\ref{tab:obs_sum}, and are detailed below in chronological order.

Spectra of PG\,0010+281 were obtained with the High Resolution Echelle Spectrometer (HIRES; \citealt{1994Vogt}) on the Keck I telescope on Mauna Kea 2013 September 17 and 20 with different setups \citep{2015Xu}. The program ID for these observations is U067Hr.  On the first night, the C5 decker was employed, which has a slit width of 1.15\,arcsec and a nominal resolving power of $R\approx36\,000$.  Two exposures were taken with integration times of 2400\,s each, using the red cross disperser (HIRESr) and covering $4960-6730$\,\AA. This resulted in a S/N of 31 in the H$\upalpha$ order. On the second night, the C1 decker was used with a slit width of 0.86\,arcsec\footnote{This value, confirmed using the Keck archive, disagrees with the setup reported in \citet{2015Xu}}.  Using this setup, two further exposures were obtained at a higher resolving power of $R\approx48\,000$, but with exposures of 1800\,s and 426\,s, with the blue cross disperser (HIRESb) covering $3050-5940$\,\AA. The average S/N for these two exposures in the H$\upbeta$ region is 39. These data were reduced using the {\sc hires redux} package\footnote{\url{https://www.ucolick.org/~xavier/HIRedux/index.html}}.

Spectroscopic observations were also carried out on the 2.5\,m Isaac Newton Telescope (INT) at Roque de los Muchachos under program P2. Fourteen spectra were obtained with grating R400V on 2015 July 22 and 23 using 420\,s exposures, and on-chip binning of 2 pixels in the spatial direction.  Six additional spectra were acquired with grating R1200R on three consecutive nights starting 2015 July 29 and using 1500\,s exposures.  Both data sets were obtained with the Intermediate Dispersion Spectrograph (IDS), a double arm, long-slit spectrograph mounted at the Cassegrain Focal station.  The REDPLUS2 detector was used in both cases, with wavelength coverage $3800-7150$\,\AA \ for R400V, and $6050-7050$\,\AA \ for R1200R.  The data were processed and extracted using {\sc molly}\footnote{{\sc molly} was written by T. R. Marsh and is available at \url{http://www.warwick.ac.uk/go/trmarsh/software}}, where typical S/N values were 20 and 15, respectively, for the R400V and R1200R grating data.

Further spectroscopic data were acquired on the Multiple Mirror Telescope (MMT), located on the summit of Mount Hopkins.  A total of four exposures were obtained on 2016 January 15 using the Blue Channel spectrograph in cross-dispersed mode with the Echellete grating \citep{1989Schmidt}.  Exposure times were set to 600\,s and cover a wavelength range from around 3100 to 7200\,\AA.  The average seeing was 1.4\,arcsec and the airmass ranged between 1.3 and 1.8 during the observations.  These data were reduced without a pipeline, using the echelle data package within {\sc iraf}. An average S/N of 20 was achieved.

PG\,0010+281 was observed with VLT UVES instrument under proposal 0100.D-0133 in service mode. A total of eight high-resolution, echelle spectra were acquired between 2017 October 22 and 2017 November 18. Observations were split into four sets of two exposures, with each set done on separate nights to reduce the probability of observing a (candidate) binary system at the same phase.  A 1.0\,arcsec slit was used resulting in $R\approx40\,000$.  The integrations were 600\,s each and $2\times2$ on-chip binning was employed. Spectra were extracted similarly to GD\,1400 as described above using {\sc esoreflex}, but with one important difference; there are numerous, residual sky emission lines\footnote{All residuals were robustly confirmed to be stationary and of telluric origin.  This shortcoming in the UVES pipeline reduction (and Phase 3 products) for this target was reported to ESO but not resolved.} in four out of the eight spectra, including one at H$\upalpha$.  To effectively minimize these residual emission lines, the sky subtraction method (the {\sf reduce.extract.skymethod} parameter) was changed from optimal to median.  In the end product S/N varied from 15 to 20 in the H$\upalpha$ region with a representative S/N of 17.

The Intermediate-dispersion Spectrograph and Imaging System (ISIS), on the 4.2\,m William Herschel Telescope (WHT), at Roque de los Muchachos was employed to obtain 13 spectra of PG\,0010+281.  A total of 11 spectra were obtained on 2018 August 29, and two further spectra on the nights of 2018 September 1 and 3 under program P8/N13.  This observing cadence provided good sensitivity to short orbital periods of several hours, as well as variations beyond one day.  ISIS has two spectrographic arms, where the blue side used the R600B grating centred at 4400\,\AA, while the red side used R1200R centred at 6560\,\AA, and where a resolving power of $R\approx7200$ was estimated.  The spectra were binned on-chip, using $3\times2$ spectral vs.\ spatial pixels in the blue arm, and $3\times1$ binning in the red arm.  The exposure time was consistently 1200\,s for both arms.  The spectra were extracted using \textsc{iraf} standard procedures.  The resulting spectra had an average S/N of 24 in the red arm.

High-resolution spectroscopy was also acquired under program 2019B-0162 with the Gemini Remote Access to CFHT ESPaDOnS Spectrograph (GRACES; \citealt{2014Chene}) located at the Gemini North Observatory on Mauna Kea.  The spectra were collected in two-fibre mode for superior sky subtraction. A total of 12 exposures were collected, consisting of four observational blocks with three exposures per block, where one block each was observed on 2019 November 10 and 11, and two blocks executed on 2019 November 12.  Individual exposures were 1800\,s, with the aim to achieve S/N $>20$ at H$\upalpha$ and $R\approx40\,000$.  The spectral data have some modest wavelength gaps, and span optical wavelengths redward of 3950\,\AA \ (including H$\upepsilon$), below which the silver mirror coating on Gemini has little reflectivity.  The Gemini spectra were reduced using the Data Reduction and Analysis for GRACES ({\sc dragraces}; \citealt{2020Chene,2021Chene}) software package.  Automated steps include overscan correction, bias subtraction, order rectification, slit tilt correction, background illumination correction, order extraction, and wavelength calibration.  The software does not include a cosmic ray removal procedure, and such artefacts were removed by identifying features not present in all three exposures in a single block.  The sky was subtracted by scaling the sky spectra by a fixed factor of 1.25 and applying a small wavelength shift (below $\pm1$\,pixel) that minimises the root mean square statistic of the resulting spectra using the \textsc{iraf} task {\sc skytweak}.

\subsubsection{Infrared spectroscopy}

To better characterise the infrared excess of PG0010+281, data from the Near-Infrared Echelle Spectrograph (NIRSPEC; \citealt{2018Martin}) were extracted from the Keck II telescope archive, and additional spectra were obtained at the NASA Infrared Telescope Facility (IRTF), both on Mauna Kea.

The NIRSPEC observations were carried out using the Keck II telescope on 2015 May 26 as part of program 15B/U049NS\footnote{These data were taken prior to the instrument upgrade in 2018 and thus some of the old configurations noted are obsolete or altered.}.  The NIRSPEC-6b filter configuration was used together with a 0.57\,arcsec slit, resulting in approximate wavelength coverage of $1.94-2.32$\,$\upmu$m \citep{2003McLean} and resolving power $R\approx2000$.  A total of 16 exposures of 300\,s each were taken, in an ABBA nodding pattern along the slit.  These data were reduced via the {\sc redspec} package \citep{2015Kim}, including a telluric correction routine that enables interpolation over real spectral features, such as the Br$\upgamma$ line in the observed telluric standard star HD\,9711.  Each AB pair of spectra were extracted and combined, resulting in eight individual spectra, and these were further combined into a single spectrum using an unweighted mean.  The resulting combined spectrum had an estimated S/N of 49.

At the IRTF, spectroscopy of PG\,0010+281 was obtained on 2019 October 20 using the SpeX near-infrared spectrograph \citep{2003Rayner}, under program 2019B030.  The spectra were taken in prism mode with a 0.5\,arcsec slit width, covering $0.80-2.52\,\upmu$m at a resolving power of $R\approx130$, obtaining 16 exposures of 120\,s each. The spectra were extracted and combined using a robust weighted mean, with telluric correction and flux calibration performed using \textsc{spextool} \citep{2004Cushing} and with HD\,222749 as an A0V standard star.  The S/N of the combined spectrum is 15.

\subsubsection{Time-series photometry}

PG\,0010+281 was observed by {\em TESS} in Sector 17 from 2019 October 8 until 2019 November 2 under designation TIC\,437743290 (\textit{TESS} Mag = 16.0\,mag; \citealt{2018Stassun}). The target was observed with spacecraft's camera number 1 and CCD chip number 2. Unlike in the case of GD\,1400, PG\,0010+281 is fainter and is surrounded by relatively bright objects, resulting in a low ratio of PG\,0010+281 flux to total flux of 0.69. The data were accessed similarly to GD\,1400 and 120\,s PDCSAP light curve was obtained. Out of 18\,012 data points, 5139 have non-zero (anomalous) quality flags, and further scrutiny showed these data to be of lesser quality and were thus rejected from analysis.

Two additional sources of photometry were retrieved to search for possible variations caused by an orbiting companion.  Publicly available infrared light curves obtained by the \textit{Wide-field Infrared Survey Explorer} (\textit{WISE}; \citealt{2010Wright}) were downloaded from NASA/IPAC Infrared Science Archive, using the AllWISE Multiepoch Photometry Table for the 3.4\,$\upmu$m \textit{W1} and 4.6\,$\upmu$m \textit{W2} bands only.  The same archive provides access to optical photometry for PG\,0010+281 within the Zwicky Transient Facility (ZTF; \citealt{2019Bellm}), where all available data were downloaded with the exception of observations with non-zero {\sf catflags} values, which were dropped.

\section{Analysis and results: GD\,1400}\label{sec:data1}

\subsection{Stellar parameters}

Stellar parameters for GD\,1400A were estimated using the Bayesian Analysis for Stellar Evolution with Nine Parameters (BASE-9; \citealt{2006vonHippel, 2013OMalley}), determined from the \textit{Gaia} DR3 parallax measurement and Pan-STARRS $griz$ photometry. The $y$-band flux was not used due to contamination from the companion.  The posterior distributions are non-Gaussian, with means and 68 percentile credible intervals $T_{\rm eff} = 11\,000\pm100$\,K, $M=0.592\pm0.009$\,M$_\odot$.  These temperature and mass estimates include formal errors that reflect only the precision within the adopted methodology.  To obtain more realistic uncertainties, the standard deviation of recently published results \citep{2011Gianninas, 2018Leggett, 2020Napiwotzki, 2021Bergeron, 2021Gentile} and BASE-9 estimates was calculated and adopted as the final uncertainty.  This results in $T_{\rm eff} = 11\,000\pm500$\,K and $M=0.59\pm0.07$\,M$_\odot$\footnote{The gravitational redshift is consistent with a mass in the lower part of this range, see Table~\ref{tab:gd1400_params} and Section~3.5}.

For radial velocity measurements of GD\,1400, the infrared and optical data were processed in a distinct manner.  The ISAAC spectra are feature-rich due to CO molecular bands from the brown dwarf companion. These individual transitions are resolved, sharp, cover almost the whole spectral range, and remain identifiable even in the lowest S/N spectrum. Therefore, cross-correlation is well-suited to calculate radial velocity changes in this case.

However, cross-correlation methods are less effective for extracting radial velocity estimates from optical spectra of white dwarf Balmer lines, which are pressure broadened and few \citep{2001Napiwotzki}.  Instead, radial velocities can be measured by fitting the relatively narrow Balmer line cores using Gaussian and Lorentzian profiles (e.g.\ \citealt{1999Maxted}). These functions individually may describe the line profile only approximately and so, for accurate modelling, a combination of multiple functions may be necessary. In general, the more pronounced core of H$\upalpha$, and to a lesser extent H$\upbeta$, can be fitted more exactly by a combination of a Lorentzian, plus a Gaussian for the wings, while higher Balmer lines can be modelled by Gaussian profiles alone. The line cores originate from the low density, non-local thermodynamic equilibrium (NLTE) regions in the upper layers of white dwarf atmospheres and thus undergo diminished pressure broadening, where H$\upalpha$ and H$\upbeta$ have the strongest NLTE features for fitting. The optical spectra of GD\,1400 (and PG\,0010+281 in Section~\ref{sec:data2} below) were fitted by minimising residuals between a combination of these functions and the spectra.

\subsection{Optical radial velocities of GD\,1400A}

The UVES spectra cover several Balmer absorption lines, but only H$\upalpha$ and H$\upbeta$ were used for radial velocity estimation of the white dwarf.  H$\upalpha$ displays a sharp NLTE core in every spectrum, while relatively narrow cores can only be distinguished in about $2/3$ of H$\upbeta$ lines. Because of this, both H$\upalpha$ and H$\upbeta$ were fitted simultaneously. To fit the lines, two dependant functions were used, each function being a combination of a Gaussian (for wider wings) and a Lorentzian (for the core). The logarithm of the wavelength was used to ensure the value of the shifts from the known air wavelength positions would be the same for both H$\upalpha$ and H$\upbeta$ during the fitting process. The fitted parameters were obtained from non-linear, least-square minimisation algorithm implemented by the \textsc{lmfit} python package \citep{2016Newville}.  The log wavelength range supplied to the fitting model varied from spectrum to spectrum, determined by first fitting H$\upalpha$ and H$\upbeta$ with a Gaussian over $\log[\uplambda($\AA$)]=3.8169-3.8173$ and $3.6866-3.6869$, respectively.  The wavelength range passed to the final fitting model would then be equal to the estimated position of the centre $\pm0.0004$\,dex. In other words, the length of the fitted range would be the same as in the first case but centred on the NLTE core from the initial Gaussian fits. The fitted wavelength range thus corresponded to approximately $\pm6$\,\AA \ for H$\upalpha$, and $\pm5$\,\AA \ for H$\upbeta$. The final wavelength central positions were then converted into velocity shifts, and barycentric corrections were applied using the \textsc{astropy} package.

\subsection{Infrared radial velocities of GD\,1400B}

Cross-correlation for radial velocity information requires a template spectrum.  For the template, it was found that using a synthetic template provided notably superior results compared to using the highest S/N spectrum or a velocity-adjusted co-add of multiple spectra.  As an initial guess, a synthetic model of a substellar, cloud-free atmosphere was picked with $T_{\rm eff} = 1500$\,K, $\log\,[g({\rm cm\,s}^{-2})]=5.5$, and solar metallicity from a grid of Sonora Bobcat models \citep{2021Marley}. The wavelength range used for cross-correlation was $2.28-2.34\,\upmu$m for both the synthetic and science spectra. The resolution of the synthetic template was reduced to the same value as the ISAAC spectra by convolution with a Gaussian kernel, and applying a flux-conserving re-sampler. To obtain the radial velocities, the \textsc{iraf} cross-correlation task {\sc fxcor} was used. 

%%%FIGURE 2%%%
\begin{figure}
\includegraphics[width=\columnwidth]{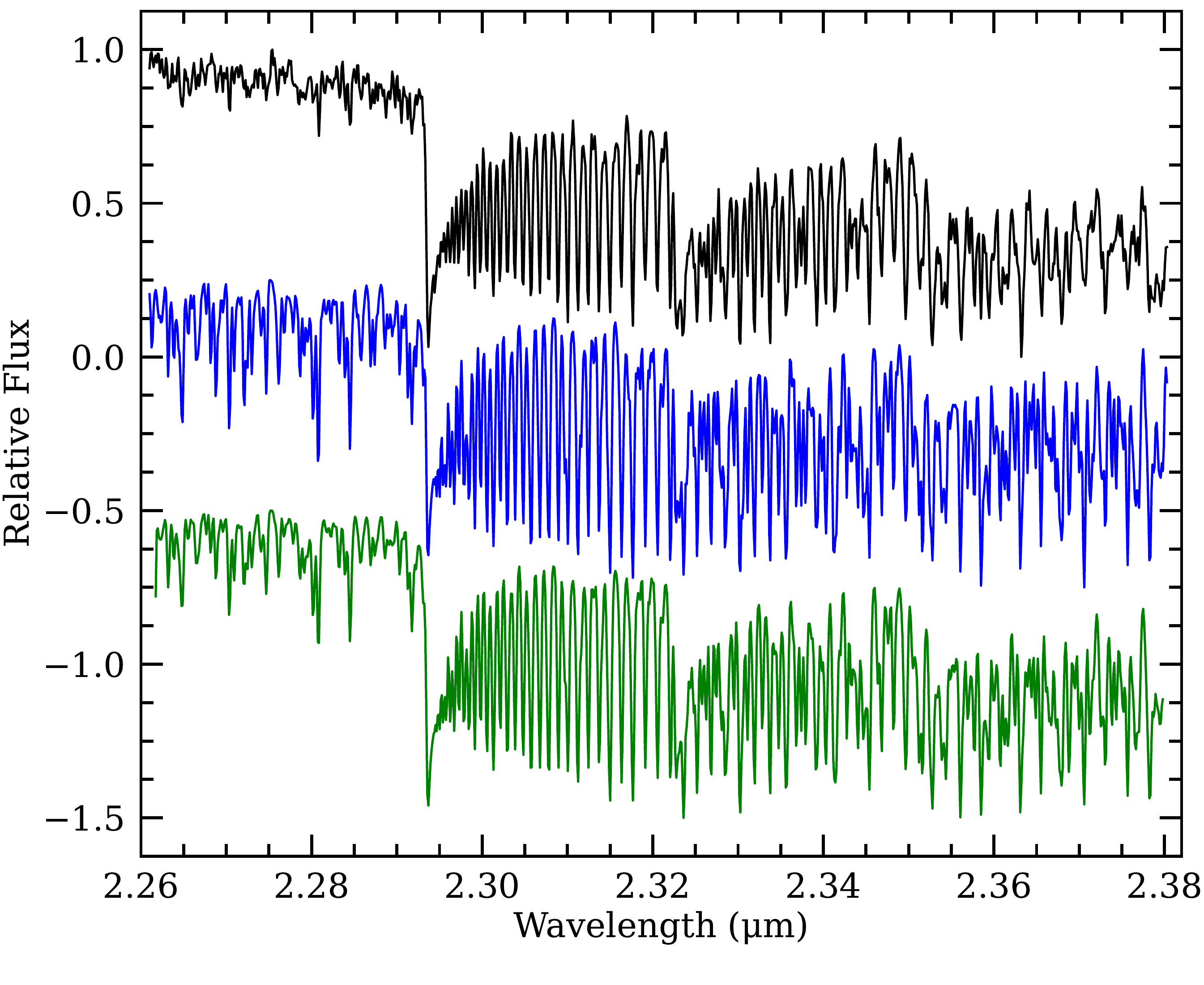}
\vskip 0mm
\caption{In black is the infrared spectrum of GD\,1400, shown as the rest-frame co-add of all ISAAC exposures, where velocity shifts were determined from the sinusoidal model fit to the radial velocities (see Figure~\ref{fig:gd1400_rvs}).  In blue is a synthetic brown dwarf template with solar metallicity, $T_{\rm eff}=1900$\,K, and $\log\,g=5.25$ from the grid of Sonora Bobcat models \citep{2021Marley}. In green is another solar metallicity template at $T_{\rm eff}=2300$\,K, and $\log\,g=5.5$ from BT-Settl model spectra \citep{2015Baraffe}.  The resolutions of both synthetic spectra were adjusted to match the $R\approx8900$ resolving power of the data.  Of all models tested, these two templates best match the ISAAC data, yet both fail to reproduce the notably weak-lined $2.26-2.29\,\upmu$m region.  At these wavelengths, spectral lines should betray only modest dilution by the white dwarf photosphere.}
\label{fig:gd1400_sonora}
\end{figure}

After obtaining velocity estimates based on the model template, barycentric velocities were calculated using the \textsc{astropy} package. The resulting velocities were then fitted with a simple sinusoidal model. Velocity shifts to the ISAAC data were then applied derived from this model to produce individual spectra that have zero velocity in the brown dwarf rest frame.  These brown dwarf rest frame spectra were then co-added into a single spectrum that was used to perform a grid search for the best fitting synthetic spectrum via residual minimisation. Two model grids were explored: the aforementioned cloud-free, Sonora Bobcat theoretical spectra, and the BT-Settl cloudy models \citep{2015Baraffe}. 

The Sonora Bobcat search grid covered $T_{\rm eff}$ from 1400 to 2400\,K in steps of 100\,K, and $\log\,g$ from 3.5 to 5.5 in steps of 0.25\,dex.  The BT-Settl grid is coarser, and was searched from $T_{\rm eff} = 1400$\,K  to 3000\,K, and from  $\log\,g=3.5$ to 5.5 in steps of 0.5\,dex.  For the Sonora Bobcat models, the search included metallicities [Z/H] = $-0.5,0,+0.5$.  However, the theoretical spectra were insufficiently distinct in the narrow region of the ISAAC spectra, and thus unable to constrain the metallicity of the brown dwarf.

The Sonora Bobcat grid search identified the best fitting model with parameters $T_{\rm eff} = 1900$\,K and $\log\,g=5.25$, where this template was then used to repeat the radial velocity cross-correlation.  The co-added, rest frame spectrum of GD\,1400B is plotted in Figure~\ref{fig:gd1400_sonora}, together with the Sonora Bobcat synthetic template that provides the best match.  It should be noted, however, that all of the templates within the model-grid search were adequate for cross-correlation with the science data, and effectively produce similar radial velocity estimates. This is further supported by the $R^2$ correlation scores, with the second best matching template ($T_{\rm eff} = 1900$\,K and $\log\,g=5.0$) having a score only 0.3 per cent lower. Similarly, the second best synthetic spectrum, at the fixed $\log\,g=5.25$ of the best match, has $T_{\rm eff} = 1800$\,K and a 1.0 per cent lower score.

A warmer result is obtained using the BT-Settl models, where the best cross-match predicts $T_{\rm eff}=2300$\,K and $\log\,g=5.5$, also shown in Figure~\ref{fig:gd1400_sonora}.  Established on the average residual error, this model provides a decent improvement of 28 per cent compared to the best fitting Sonora Bobcat template, and notably predicts weaker lines in the $2.26-2.29\,\upmu$m region.  Nevertheless, both sets of atmospheric models exhibit lines that are visibly too strong compared to the weak lines observed at the blue end of the spectra.  The results from both models were averaged, yielding $T_{\rm eff}=2100\pm200$\,K and $\log\,g$ in the range $5.25-5.5$. 

\subsection{Orbital period determination}

The bulk of UVES spectra for GD\,1400A were taken sparsely over a period of three months, yielding an orbital period of $9.978\pm0.002$\,h.  Based on the initial determination of this period with UVES, the ISAAC data for GD\,1400B were obtained more efficiently, using continuous observations over two consecutive nights, resulting in a period of $9.99\pm0.01$\,h.  Both independent results were obtained from a mean value of the most significant peaks within 1000 Lomb-Scargle (LS) periodograms \citep{Lomb1976,Scargle1982} run on bootstrapped samples. The uncertainties quoted are two standard deviations of the set of most significant peaks.  Another period inference was made using both the UVES and ISAAC data as follows.  The radial velocity estimates from each instrument were fitted individually with a sine curve, resulting in a semi-amplitude and period for each binary component.  The radial velocities were then offset by the fitted $\upgamma$ value, and divided by their respective, fitted velocity semi-amplitudes for normalization (ISAAC data were divided by $-K_2$ because of the anti-phase relationship between the white dwarf and brown dwarf velocities). Then the two normalized datasets were combined and another period was calculated, again using the bootstrapping method, but this time taking into account uncertainties associated with each instrument (for ISAAC data the measurement weight is approximately $4.5\times$ greater than UVES). This combined analysis results in $P=9.976\pm0.006$\,h.  All three period values are consistent, and the UVES estimate of $9.978\pm0.002$\,h is adopted as the final value, on the basis of the highest precision.  This period was then used for all subsequent analyses, namely to re-fit the radial velocity curves for each instrument individually, to obtain the final amplitude and $\upgamma$ estimates while keeping the period fixed.  

%%%FIGURE 3%%%
\begin{figure}
\includegraphics[width=\columnwidth]{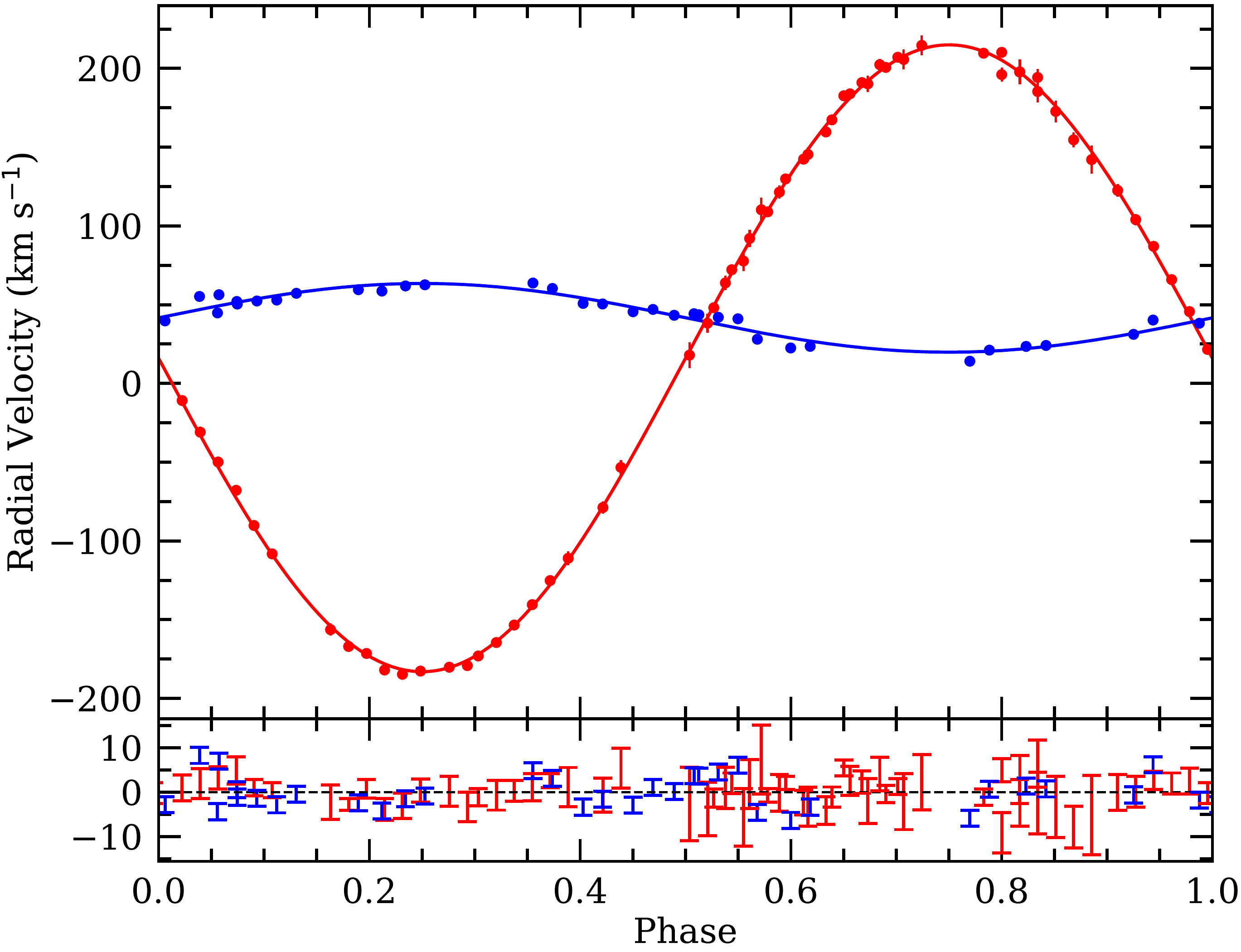}
\vskip 0mm
\caption{The upper panel plots the radial velocity phase curves of GD\,1400A and GD\,1400B in blue and red, respectively. The measurements for both components are well described by sine functions that have been fitted to the data (circular orbits are expected post-common envelope).  The lower panel plots the residuals between the measurements and the simple model.}
\label{fig:gd1400_rvs}
\end{figure}

The radial velocity curves for both binary components of GD\,1400 are shown in Figure~\ref{fig:gd1400_rvs} and the derived orbital and binary parameters are summarised in Table~\ref{tab:gd1400_params}.  Of particular note is the impressive performance of the cross-correlation on the infrared spectra, where the final uncertainty on $K_2$ is less than 1\,km\,s$^{-1}$ (cf.\ the instrumental resolution 33\,km\,s$^{-1}$).  This can further be appreciated in Figure~\ref{fig:gd1400_sonora}, where the match between the synthetic templates and the empirical spectrum of CO bandheads is striking.

%%%TABLE 2%%%
\begin{table}
\centering
\caption{Orbital and (sub)stellar parameters of the GD\,1400 binary.}
\label{tab:gd1400_params}
\begin{tabular}{rl} % four columns, alignment for each

\hline

$P$ (h)                                     &$9.978\pm0.002$\\
$T_0$ (MJD)                                 &$55498.389\pm0.002^a$\\
$K_1$ (km\,s$^{-1}$)                        &$21.8\pm1.1$\\
$K_2$  (km\,s$^{-1}$)                       &$199.2\pm0.6$\\
$\upgamma_1$ (km\,s$^{-1}$)                 &$41.7\pm0.7$\\
$\upgamma_2$ (km\,s$^{-1}$)                 &$16.0\pm0.4$\\
$M_1$ (M$_\odot$)                           &$0.59\pm0.07^b$\\
$M_2$ (M$_{\rm Jup}$)                       &$68\pm8$\\
$i$ (\degr)                                 &$60\pm10$\\
$|\upgamma_1 - \upgamma_2|$ (km\,s$^{-1}$)  &$25.7\pm0.8$\\
$GM_1/R_1c$ (km\,s$^{-1}$)                  &$28.9\pm4.0$\\

\hline

\end{tabular}

\flushleft
$^a$ $T_0$ is provided for phase $= 0$ (see Figure~\ref{fig:gd1400_rvs}).\\
$^b$ $M_1$ derived from photometry and parallax is in mild tension with the gravitational redshift implied by $|\upgamma_1 - \upgamma_2|$ (see Section~3.5).
\end{table}

The mass of GD\,1400B is calculated from Kepler's law in the form of $M_2 = M_1 (K_1/K_2) = 68\pm8$\,M$_{\rm Jup}$ ($0.065\pm0.008$\,M$_{\odot}$), which places it at the more massive end of substellar objects, not just those that are companions to white dwarfs.  The BASE-9 estimate of $M_1$ with formal errors yields $M_2=68\pm4$\,M$_{\rm Jup}$, and is modestly below the hydrogen-burning limit of approximately 75\,M$_{\rm Jup}$ \citep{1997Burrows} for solar metallicity.  For a circular orbit, the semimajor axis is 0.009\,AU or about 1.9\,R$_\odot$, and is outside of the distance required for the substellar companion to fill its Roche lobe.  The difference between the measured binary component velocities, $\upgamma_1$ and $\upgamma_2$, is $25.7\pm0.8$\,km\,s$^{-1}$, and is in agreement with the lower range of gravitational redshifts $28.9\pm4.0$\,km\,s$^{-1}$ expected for a $0.59\pm0.07$\,M$_\odot$ white dwarf.  At face value, the measured $|\upgamma_1-\upgamma_2|$ corresponds to a white dwarf mass of 0.54\,M$_\odot$, and {\em suggests the actual masses of GD\,1400A and B are towards the lower ends of the ranges of values allowed by their uncertainties.}

\subsection{Discussion}

Despite being a well-known pulsator \citep{2003Fontaine} with numerous frequencies reported using Sector 3 \textit{TESS} observations \citep{2020Bognar}, any day-night side or irradiation effect, if present, should be detectable in time-series photometry.  However, in two sectors of {\em TESS} data, which are by far the most sensitive photometric data available, there is no signal present at the orbital period of the binary.  Neither is there any indication of photospheric heating (e.g.\ emission lines) in the day-side co-add of ISAAC spectra for GD\,1400B, nor any notable differences in the day- and night-side co-added spectra.  It may be that the 10\,h orbital period (to which the brown dwarf should be tidally locked) and the modest $T_{\rm eff}$ of the primary are simply insufficient to give rise to an observable modulation.  Lastly, a co-add of UVES spectra in the brown dwarf rest frame also failed to reveal any emission lines that could be associated with irradiation, thus strengthening the idea that surface heating is mild or absent at all wavelengths considered here.  

A similar lack of observed irradiation was reported for ZTF\,J0038+2030 \citep{2021vanRoestel}, a 10\,900\,K white dwarf with a brown dwarf companion in a 10.4\,h orbit.  GD\,1400 is $10\times$ brighter than ZTF\,J0038+2030, and thus the spectroscopic data as well as {\em TESS} photometry, are more sensitive.  These two white dwarfs have similar effective temperatures, but the radius of GD\,1400A is around 10 per cent smaller, owing to a mass that is 15 per cent higher, and the expected temperature increase on the surface of GD\,1400B is likely no more than 10\,K \citep{2014Littlefair}.  Thus, the lack of any observed photometric modulation or emission lines is consistent with expectations.

Of additional interest, given that GD\,1400 is a ZZ\,Ceti, stellar rotation may split the pulsation frequencies into 2$\ell+1$ azimuthal components. Such equally-spaced triplets in a periodogram, provided they are association with the dipole ($\ell = 1$) mode, could help uncover the rotation period of the white dwarf.  No such triplets or quintuplets (for the case of $\ell = 2$) could be confidently identified in either the 120 and 20\,s {\em TESS} data.  However, the white dwarf may not be tidally locked (unlike the brown dwarf), thus the orbital period may not equate to the white dwarf rotation rate.  

The brown dwarf in this system is an excellent benchmark for existing models, where the temperature, mass, and age of GD\,1400B can all be estimated.  Although the system experienced a common envelope, the mass of the white dwarf suggests that single star evolution was not drastically truncated, and provides a lower limit on the system age via white dwarf cooling at around 0.5\,Gyr.  The white dwarf mass is consistent with a CO-core remnant that completed the RGB and where the common envelope was initiated further out during the AGB.  However, the current white dwarf mass is almost certainly lower than would have resulted for a single star, and thus a total age for the binary system, using initial-to-final mass relations (IFMRs) and main-sequence lifetimes, will be over-predicted and unreliable.  Independent from evolutionary considerations of the primary, the 3D space motion of GD\,1400 is $(U,V,W)=(3,-1,-6)$\,km\,s$^{-1}$, and consistent with a relatively young, thin disk star with a likely age less than around 2\,Gyr. 

Based on brown dwarf evolutionary models, at an age of 1.0\,Gyr, a 68\,M$_{\rm Jup}$ object should have an effective temperature near 1800\,K and surface gravity $\log\,g\approx5.3$ \citep{2015Baraffe}.  Of the spectral templates tested above, a Sonora Bobcat model with $T_{\rm eff}= 1800$\,K, $\log\,g=5.25$ \citep{2021Marley} provides an adequate fit to the ISAAC spectrum over the CO bandheads.  However, if GD\,1400B has $T_{\rm eff}\approx2100$\,K, as inferred by considering both cloudy and cloud-free model fits to the $K$-band ISAAC spectrum, then it has a brown dwarf cooling age of only 0.6\,Gyr.  If the actual substellar companion mass is within the lower range of values allowed by the uncertainties, as suggested by the gravitational redshift for $M_1$, then its cooling age will be somewhat younger.

The white dwarf cooling age is $0.48\pm0.01$\,Gyr and can thus accommodate a total system age less than 1\,Gyr via binary evolution, consistent with the relatively warm brown dwarf temperature implied by the theoretical model fits.  An alternative history is that the total age is somewhat older, but that GD\,1400B has been re-heated during circularization and tidal synchronization.  This possibility has yet to be explored from a theoretical perspective, and is beyond the scope of this work, but a potentially useful speculation because at least one substellar companion to a white dwarf appears to be unexpectedly inflated (i.e.\ has a radius too large for its evolutionary age; \citealt{2020Casewell_a}).

It should be emphasized that the $T_{\rm eff}$ for GD\,1400B has been derived in an unconventional way, by fitting spectral templates to a $K$-band spectrum covering a narrow wavelength range, and at moderately high spectral resolution.  Essentially, only the CO bands have been fitted, and while these features are distinctive, providing excellent anchor points for model comparison, this atypical determination of brown dwarf parameters may result in systematic errors.  To underscore this possibility, as pointed out earlier, Figure \ref{fig:gd1400_sonora} demonstrates that neither of the synthetic spectral grids can provide a decent match to the absorption lines in the $2.26-2.29\,\upmu$m region of the ISAAC data.

The cause of this line strength mismatch between models and observations of GD\,1400B is unclear. First, it cannot be caused by dilution from the light of the primary.  The substellar companion is comparably bright at these wavelengths, and any significant dilution would similarly weaken the depths of its CO features, which is not observed.  Second, and ignoring the young disk kinematics and the likely youth of the warm brown dwarf, the weak lines cannot be due to metal poverty.  Owing to increased collision-induced absorption in the $K$-band, metal-poor L subdwarfs, and even warmer M subdwarfs, display spectra where the CO bands are typically absent or notably weaker than their solar metallicity counterparts \citep{2004Burgasser,2009Burgasser,2017Zhang,2018Zhang}.  Third, the significantly weaker lines in the observations are unlikely to be due to clouds.  The BT-Settl models exhibit moderately weaker absorption lines than the cloud-free Sonora Bobcat grid, and thus provide a better overall match to the data, but clouds are not a known source of line obscuration \citep{2015Baraffe}.  

Remaining possibilities include a particularly dusty atmosphere or current modelling deficits.  
In at least some cases, atomic line profiles in brown dwarfs can be significantly affected by dust, but dusty spectral models are currently limited, available only for a few optical lines \citep{2008Johnas}.  Comparing contemporary models with data covering a broader range of wavelengths may help shed light on this apparent discrepancy in line depths for GD\,1400B.  While there are no other L dwarfs with $R\approx9000$ $K$-band spectra, available low- to medium resolution spectra reveal few or no features in the region blue-ward of the first CO bandhead \citep{2003McLean,2005Cushing,2012Rojas}, in contrast to the strong and feature-rich models in Figure~\ref{fig:gd1400_sonora}.  Thus, the spectrum of GD\,1400B may not be unusual, but the models which remain at odds with observed L dwarf spectra.

%%%FIGURE 4%%%
\begin{figure*}
\includegraphics[width=\linewidth]{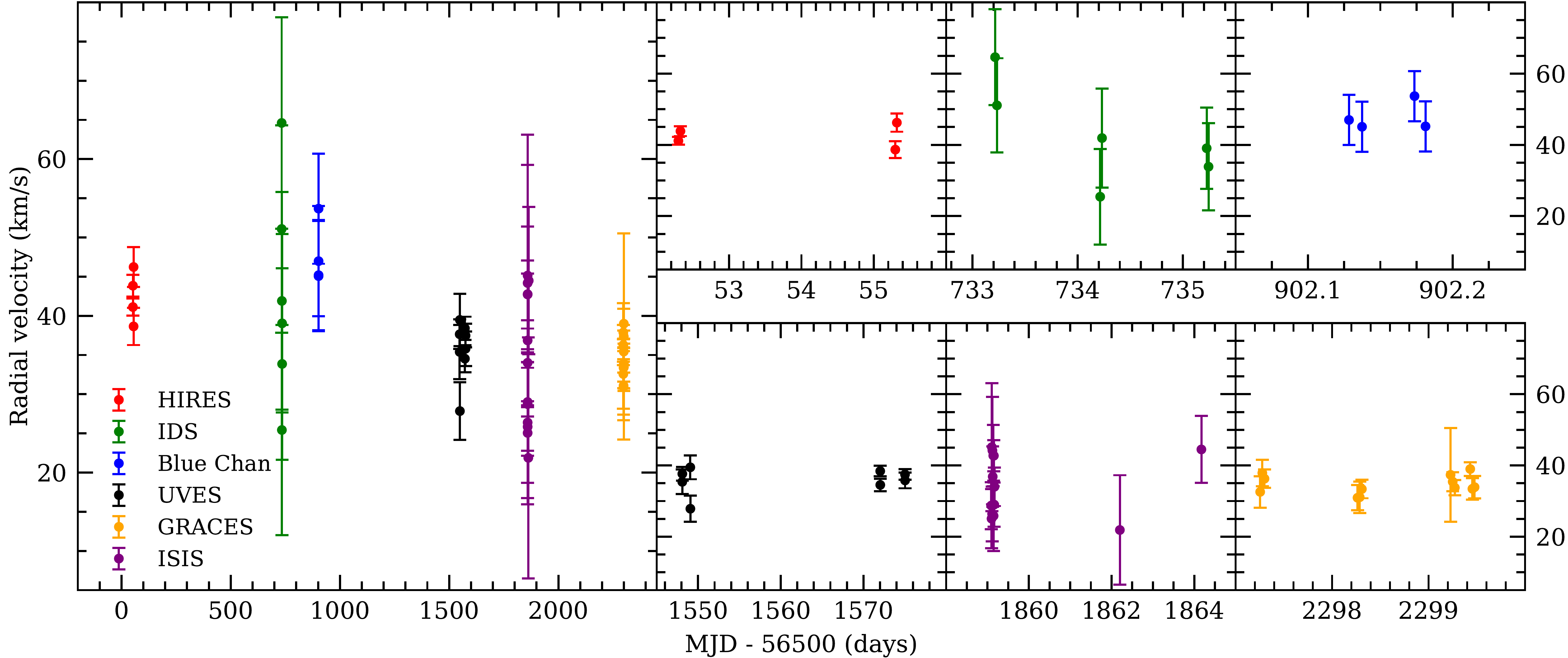}
\vskip 0mm
\caption{All radial velocities and their errors for PG\,0010+281 as determined from H$\upalpha$ line fitting.  The six panels on the right show zoomed-in regions of the left panel. The observations span just over six years, and have been acquired with six different instruments, each designated in the plot legend with a distinct symbol colour (see Table\,\ref{tab:obs_sum}).  Measurements are plotted as a function of MJD both for the overall, multi-instruments dataset, as well as for individual instruments.  The plotted data are uncorrected for any systematic offsets between instruments.}
\label{fig:pg0010_rvs}
\end{figure*}

\section{Analysis and results: PG\,0010+281}\label{sec:data2}

\subsection{Stellar parameters}

Physical parameters and ages were constrained from \textit{Gaia} DR3 parallaxes and Pan-STARRS photometry using BASE-9.  As for GD\,1400, the posterior distributions are non-Gaussian with means and 68 percentile credible intervals yielding $T_{\rm eff} = 26\,500\pm1000$\,K, $M=0.569\pm0.005$\,M$_\odot$, and a total age $7.5\pm2.5$\,Gyr assuming isolated stellar evolution.  Similarly to GD\,1400, a standard deviation of recently published results \citep{2005Liebert,2011Gianninas,2020Xu,2021Gentile} and the BASE-9 estimate modestly increases the uncertainties to $T_{\rm eff} = 26\,500\pm1100$\,K, $M=0.569\pm0.027$\,M$_\odot$. These are the adopted mass and effective temperature for this work.

The kinematics of PG\,0010+281 do not support such an old age as suggested by single star evolution.  Using {\em Gaia} DR3 astrometry, together with an estimated radial velocity from the weighted mean radial velocity (36.1\,km\,s$^{-1}$), corrected for the gravitational redshift (24.7\,km\,s$^{-1}$) of a $M=0.569$\,M$_\odot$ white dwarf, yields three dimensional space velocities $(U,V,W) = (28,1,-10)$\,km\,s$^{-1}$.  In accordance with chemo-kinematical models of the local solar neighbourhood, the white dwarf is $80\times$ more likely to be a thin disk star than a member of the thick disk \citep{2014Bensby}, and thus likely to have an age below 5\,Gyr.  Kinematics are not conclusive by any means, as any star of any age can have any space velocity, where Galactic kinematical populations are established primarily by their dispersions, and to a much lesser degree their mean space velocities.  However, the mismatch between kinematics, mass and total age suggests the possibility of binary evolution.

Because of this unexpectedly low mass and total age conflict, three further age and mass determinations were considered, two of which are fully independent.  First, instead of Pan-STARRS $grizy$ fluxes, SDSS $ugriz$ photometry was used as input for BASE-9.  This yields an even lower white dwarf mass, with a most probable total age of $9.8\pm1.9$\,Gyr, and was therefore discounted in favour of the results using Pan-STARRS photometry.  Second, the {\em Gaia} eDR3 catalogue of white dwarfs gives $M=0.52\pm0.02$\,M$_\odot$ calculated on a combination of parallax and {\em Gaia} colours, which are then fitted using a grid of atmospheric models using synthetic photometry \citep{2021Gentile}.  From published IFMRs (e.g.\ \citealt{2018Cummings}, the same as used for the BASE-9 calculations), such a low white dwarf mass would yield a progenitor that should remain on the main sequence over a Hubble time, and is thus unphysical, instead requiring truncated (binary) stellar evolution.  Third, there is a spectroscopic determination of $T_{\rm eff}$ and $\log\,g$ proceeded from Balmer line fitting, which yields $M=0.57\pm0.02$\,M$_\odot$ \citep{2011Gianninas}, in excellent agreement with the BASE-9 results based on Pan-STARRS.  Thus, all available mass estimates for PG\,0010+281 are sufficiently low that binary evolution is a possibility.

\subsection{Radial velocities}

The radial velocities of PG\,0010+281 were determined as described for GD\,1400A in Sections~3.1 and 3.2, and are plotted as a whole and centred on individual instruments in Figure~\ref{fig:pg0010_rvs}.

Uncertainties on the radial velocities of this target were estimated using the bootstrapping method. In detail, a number of points corresponding to the total number of points in a spectrum are selected at random with replacement, and the resulting Balmer profiles are refitted 1000 times per spectrum. The standard deviation of the resulting radial velocities was then estimated. In general, error estimation from bootstrapping provides a more robust result if there are imperfections in the input data, but the method performed less well for low resolution spectra. The NLTE core of H$\upalpha$ in a low resolution spectrum would consist of only a few points which can be difficult to fit if any of these points were not selected by the algorithm.

To quantify the presence of any variability or excessive scatter in the radial velocities, beyond what would be expected from a constant radial velocity, the $\upchi^2$ statistic and the resulting probability are calculated \citep{2000Maxted}. Briefly, this includes calculating a weighted mean radial velocity, then performing a goodness-of-fit test between the observed radial velocities and the calculated mean. If the constant mean value provides a poor fit to a potentially varying signal, then the signal is judged as variable.  This is quantified by calculating the probability $p$ of observing such a signal from random radial velocity fluctuations where low $p$ value signifies high scatter and thus probable variability.  The individual instrument means and resulting $p-$values are given in Table~\ref{tab:pg0010_rvsum}.

The calculated, weighted mean radial velocities for the individual spectrographs can identify systematic offsets between the sets of observations. To ascertain whether or not such observed velocity offsets can be corrected, individual sky line velocities in raw spectra were measured.  Depending on availability, one or both of the sky lines at 5577.34\,\AA \ and 6300.30\,\AA \ were fitted, and any velocity shifts were calculated, based on their expected air wavelengths.  In the case of the MMT Blue Channel observations, and despite the fact that arc lamps were taken either immediately before or after each exposure, the sky lines varied notably between frames and the measured velocity shifts were poorly behaved.  In the case of the HIRES data, significant positive shifts were detected in all exposures; around 5 and 10\,km\,s$^{-1}$ for the HIRESr and HIRESb observations, respectively.  However, it should be noted that wavelength calibration frames were not taken concurrently with science frames, and thus the effects of telescope flexure cannot be corrected.

%%%TABLE 3%%%
\begin{table}
\centering
\caption{Summary of radial velocity measurements for PG\,0010+281.
\label{tab:pg0010_rvsum}}
\begin{tabular}{ccccc} % four columns, alignment for each

\hline
Instrument      &Year   &$v_{\rm rad}$      &$N_{\rm obs}$  &$p-$value\\
                &       &(km\,s$^{-1}$)     &               &\\

\hline

HIRES$^{\rm a}$ &2013   &$42.1\pm0.8$   &4              &0.068\\
IDS             &2015   &$42.2\pm5.2$   &6              &0.378\\
Blue Channel       &2016   &$47.7\pm3.5$   &4              &0.801\\
UVES            &2017   &$36.6\pm0.7$   &8              &0.177\\
ISIS            &2018   &$34.8\pm4.2$   &13             &0.996\\
GRACES          &2019   &$35.1\pm0.8$   &12             &0.646\\

\hline

\end{tabular}

\flushleft
$^{\rm a}$ Wavelength calibration and science frames not taken concurrently.
\smallskip

{\em Notes}.  The $p-$ value gives the likelihood the measurements are consistent with a single star with a constant radial velocity given by the weighted mean.  A value $p<0.0027$ indicates that the star is binary at $3\upsigma$ confidence.

\end{table}

\subsection{Companion limits}

Owing to significant instrumental offsets in velocities of skylines, as described above, the Blue Channel and HIRES observations were analysed individually, and were not combined with data from the other instruments.  In contrast, and using the velocity offsets derived from sky lines, the data from IDS, UVES, ISIS, and GRACES all have an average systematic offset of $\lesssim1$\,km\,s$^{-1}$.

\begin{figure}
\includegraphics[width=\columnwidth]{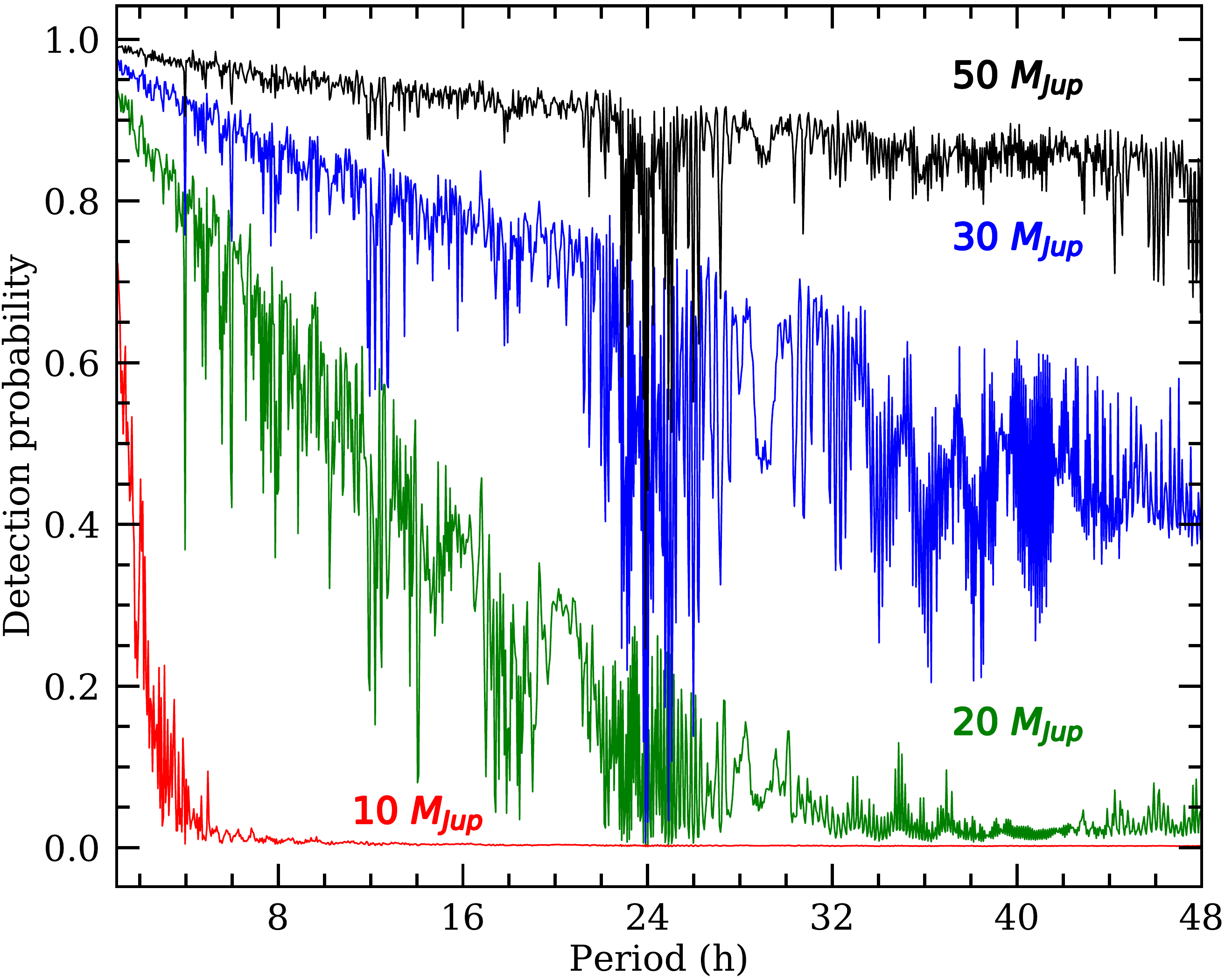}
\vskip 0mm
\caption{Detection probabilities as a function of orbital period for a range of substellar companion masses, based on a $\upchi^2$ test and the combined radial velocity data from IDS, UVES, ISIS, and GRACES.  Of particular interest is the region below 12\,h, which spans the orbital periods of all known white dwarf - brown dwarf binaries \citep{2022Zorotovic}.  These brown dwarf companions to white dwarfs have masses in the range $50-70$\,M$_{\rm Jup}$, for which the detection probabilities are typically above 0.9.}
\label{fig:pg0010_dprobs}
\end{figure}

To quantify the sensitivity of the observations to potential orbital periods, a number of simulations were ran that take into account the calculated physical parameters of PG\,0010+281 and a simulated companion of a specified mass. The simulated signal is a sine curve with a fixed period, random inclination (sampled uniformly from $\arccos(i)$ function) and random orbital phase. The resulting signal is sampled using the time stamps of the actual observations. Time stamps from IDS, UVES, ISIS, and GRACES were used in the simulations since these instruments do not show any significant systematic offsets ($\leq1$\,km\,s$^{-1}$) as determined from the sky line measurements. To assess if such a signal would be detected, a $\upchi^2$ goodness-of-fit statistical test is performed to identify whether the generated signal is statistically similar to a constant radial velocity with Gaussian uncertainties determined from the actual radial velocity measurement errors.  A generated signal is deemed recoverable or significantly variable if its $p-$value $<0.0027$ (3$\upsigma$ in a normal distribution). For each specified period, 100\,000 simulations were performed, consisting of 100 random phases for 1000 randomly sampled inclinations.  Periods from 1\,h to 2\,d were sampled in step sizes just under 3\,min, and the resulting sensitivities are plotted in Figure~\ref{fig:pg0010_dprobs}.

%%%FIGURE 5%%%

%%%FIGURE 6%%%
\begin{figure}
\includegraphics[width=\columnwidth]{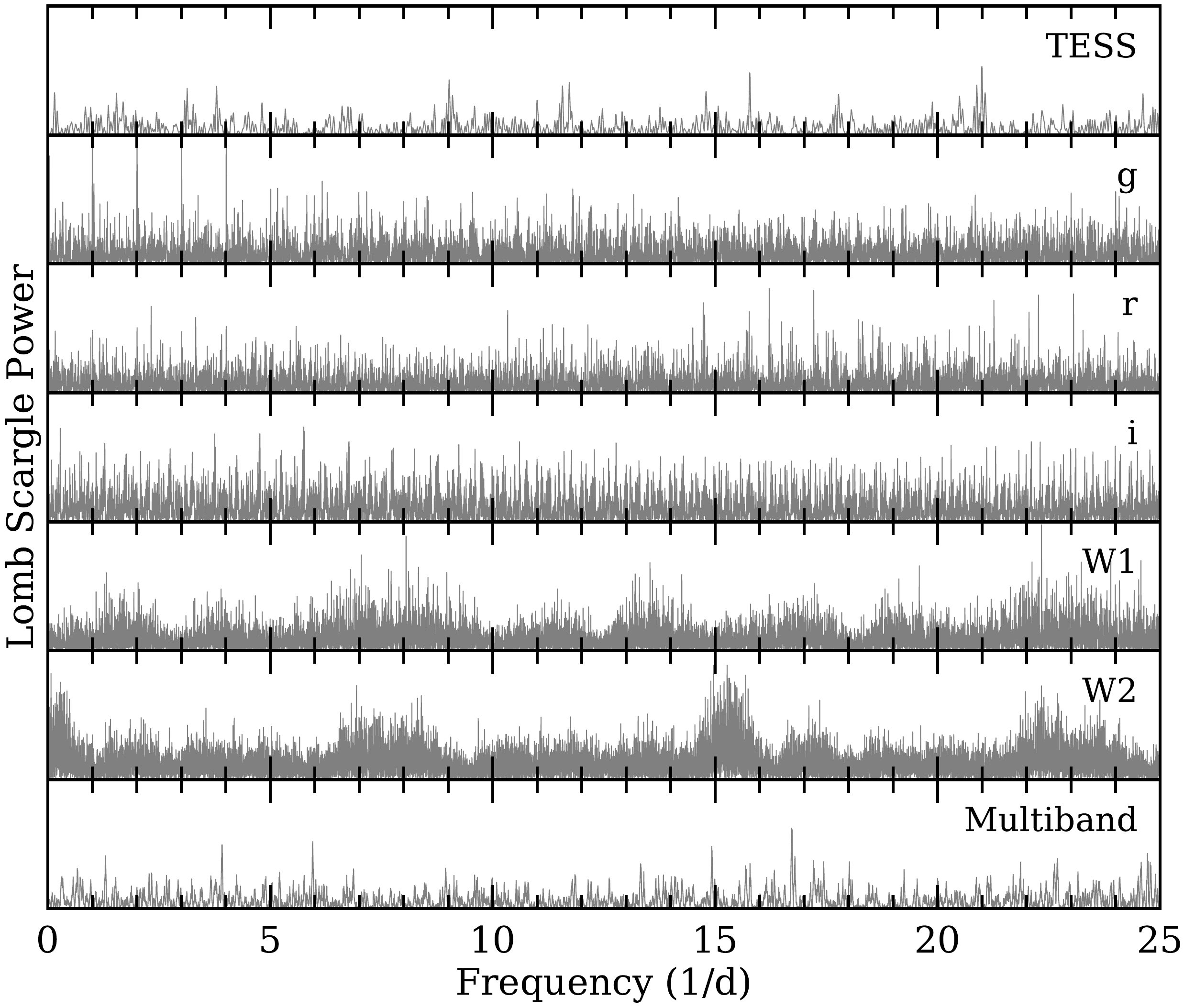}
\vskip 0mm
\caption{Lomb-Scargle periodograms for PG\,0010$+$281 photometry.  From top to bottom the panels correspond to: \textit{TESS}, ZTF $gri$, {\em WISE} $W1$ and $W2$, and a multi-band periodogram that combines all of the aforementioned photometry.  There are no significant periods except at aliases expected from ground-based observations, where the top of the y-axis range corresponds to a 1 per cent false alarm probability level in each panel.}
\label{fig:pg0010_lsps}
\end{figure}

The radial velocity observations provide excellent sensitivity to orbital periods within 12\,h.  Observations by UVES and GRACES are the most powerful, and although if treated separately there are some gaps in the sensitivity at certain periods, by combining these observations together such gaps become practically non-existent.  The small systematic offsets of IDS, UVES, ISIS, and GRACES should allow these datasets to be combined as if from a single instrument (albeit with different errors).  Overall, it can be concluded that the temporal sensitivity of the data is adequate for substellar companions that experienced common envelope evolution.

A slight and superficial long-term trend is present in the radial velocity, which is reflected in the mean radial velocity values for each instrument.  As can be seen in Table~\ref{tab:obs_sum}, the first three datasets were acquired from 2013 to 2016, while the later three were taken from 2017 to 2019.  The higher mean values from 2013 to 2016 followed by lower values from 2017 to 2019 might be construed to signify an orbital period of the order of years.  However, such an interpretation is unlikely for two reasons.  First, the sky line analysis showed that the radial velocities from HIRES and Blue Channel are less reliable. Moreover, the actual systematic offsets cannot be fully pinned down just from the sky line measurements, and would require radial-velocity standard observations, which were not acquired.  Second, unlike radial velocity data, astrometric wobbles are relatively insensitive to orbital inclination, and periods on timescales near a year should be revealed by {\em Gaia}.  However, the DR3 renormalised unit weight error is 1.044 \citep{GaiaDR3}, indicating that a single-star model provides a good fit to the observed precision astrometry.  Long term variability cannot be ruled out but appears unlikely.

%%%FIGURE 7%%%
\begin{figure*}
\includegraphics[width=\linewidth]{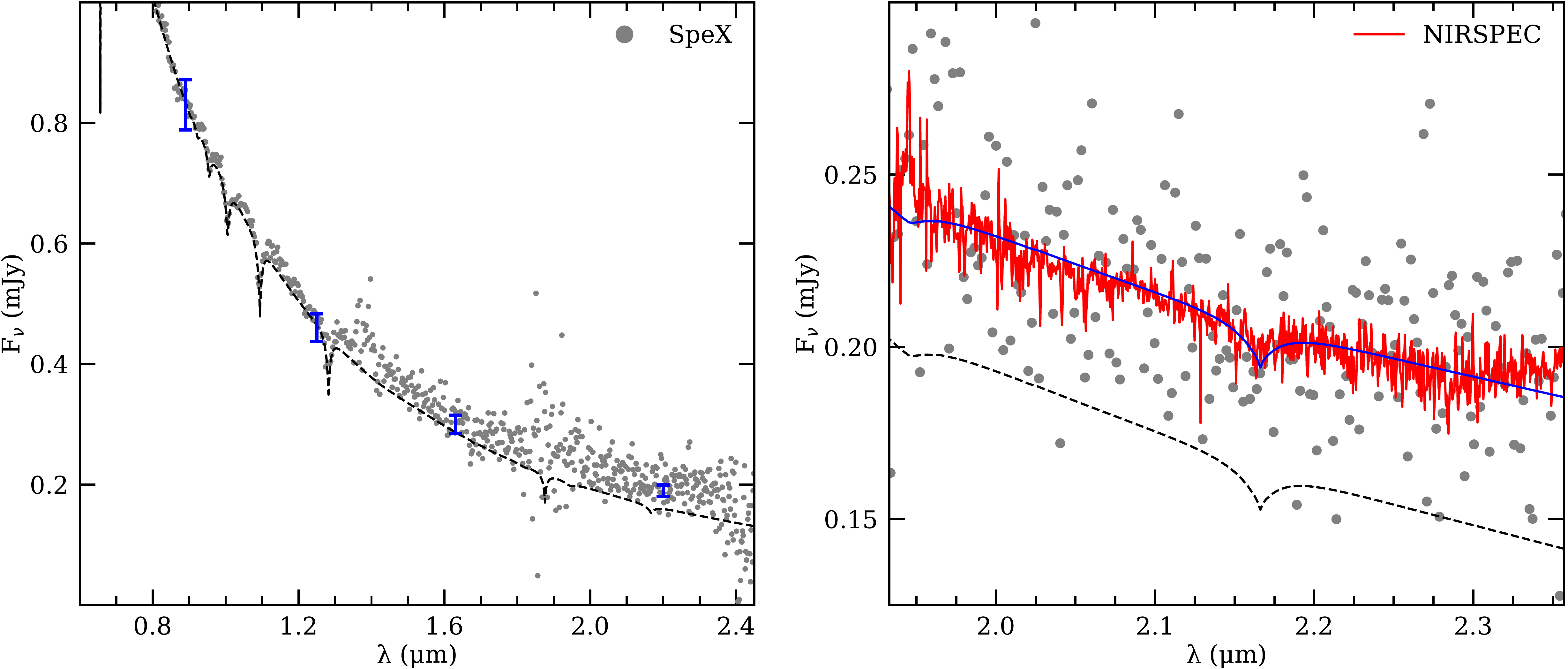}
\vskip 0mm
\caption{Infrared spectra of PG\,0010+281.  The left panel shows the SpeX prism spectrum (grey points), together with the same photometry (blue error bars) and white dwarf atmospheric model (black dashed line) as shown in Figure~\ref{fig:pg0010_irsed}.  The right panel shows the NIRSPEC spectrum in red, offset to match the $K$-band photometric flux, and with a weak but clear detection of Br$\upgamma$ from the stellar photosphere.  The SpeX data and the white dwarf model atmosphere are also plotted as on the left, where the blue solid line includes the addition of a 1000\,K blackbody.  The upturn near 2.3\,$\upmu$m in the NIRSPEC data are not seen in the SpeX data, and thus may be an artefact of the reduction process or instrument settings.}
\label{fig:pg0010_irspec}
\end{figure*}

\subsection{Time-series photometry}

The available photometric monitoring of PG\,0010+281 with \textit{TESS}, ZTF, and \textit{WISE} were analysed using the \textsc{astropy} package implementation of the Lomb-Scargle periodogram.  These results are plotted in Figure~\ref{fig:pg0010_lsps}, including a single collective light curve analysis, using the multiband implementation of the periodogram \citep{2015VanderPlas}.  As can be seen in the plot, no significant frequencies have been identified at any wavelength, and where {\em TESS} is by far the most sensitive in terms of cadence.  The lack of statistical significance was determined using a standard false alarm probability \citep{2018VanderPlas}, estimated via a bootstrapping method. The analysis shows that there are no frequencies in any of the surveys or their combination with a false alarm probability below 1 per cent, except the aliases expected from the ground-based observations in ZTF $g$ band.

While the radial velocity analysis is sensitive to the secondary mass, a heating or day-night side effect is primarily driven by the size and temperature of any irradiated spot.  The radii of low-mass stars, brown dwarfs, and giant planets are all roughly equal down to M$_{\rm Jup}$, and thus the heating effect might still be pronounced for a wide range of substellar secondary masses.  Furthermore, the white dwarf has $T_{\rm eff}\approx26\,500$\,K which is more than sufficient to drive such a variation for a range of short-period orbits and companion properties.  For a brown dwarf temperature near 1700\,K, and orbital periods within 1\,d, irradiation should raise the surface temperature by at least 100\,K, and increasing to over 200\,K for periods within 12\,h \citep{2014Littlefair}.  The fact that no photometric modulation is observed is in agreement with the radial velocity data, and there is likely no substellar companion consistent with post-common envelope evolution.

As with radial velocities, an orbital inclination that is face-on will not produce any variations from a day-night side effect, and such a signal will diminish with increasing inclination angle.  However, this requires a finely-tuned orientation, and while it cannot be ruled out, this is unlikely.

\subsection{Infrared flux excess}

In Figure~\ref{fig:pg0010_irspec} are shown the two infrared spectra taken of PG\,0010+281.  The SpeX data are consistent with a modest excess in the $H$ and $K$ bands, but are otherwise too noisy to discern any possible features.  The higher resolution and higher S/N spectrum taken with NIRSPEC exhibit two features that are possibly noteworthy.  The first is the upturn near 2.3\,$\upmu$m, which is rather abrupt and angular in the data.  While the SpeX data are noisy, the data extend to 2.55\,$\upmu$m and there is no upturn evident in that spectrum.  At present, a reduction or instrumental artefact cannot be ruled out.  The second feature is the weak but clear detection of Br$\upgamma$, consistent with a partial detection of the white dwarf photosphere, possibly diluted by another source, such as a thermal continuum from warm dust as modelled and plotted.  Overall, the infrared spectra fail to reveal any significant slope or features associated with a companion (e.g.\ \citealt{2020Casewell_b}) or a gradual upturn associated with the rise of thermal dust emission (e.g.\ \citealt{2006Kilic}), but are far from conclusive based on the available combinations of wavelength coverage and S/N. 

\subsubsection{A substellar survivor?}

The constraints from the previous Sections~4.2 and 4.4 have direct bearing on one of the hypotheses for the excess infrared emission observed with {\em Spitzer} IRAC \citep{2015Xu}.  An irradiated companion would necessarily be in a close orbit, which should most often result in radial velocity variations, emission lines and photometric variation as the heated side changes viewing configuration, especially from a relatively hot white dwarf such as PG\,0010+281.  While both radial velocity changes and photometric variations should be zero for a binary inclined by 90\degr, emission lines from a low-mass companion might still be visible at all orbital phases.  Despite this fact, there is no evidence for emission lines in any of the numerous spectra obtained for PG\,0010+281 (see Section~\ref{sec:obs}, and UVES footnote).

\begin{figure}
\includegraphics[width=\linewidth]{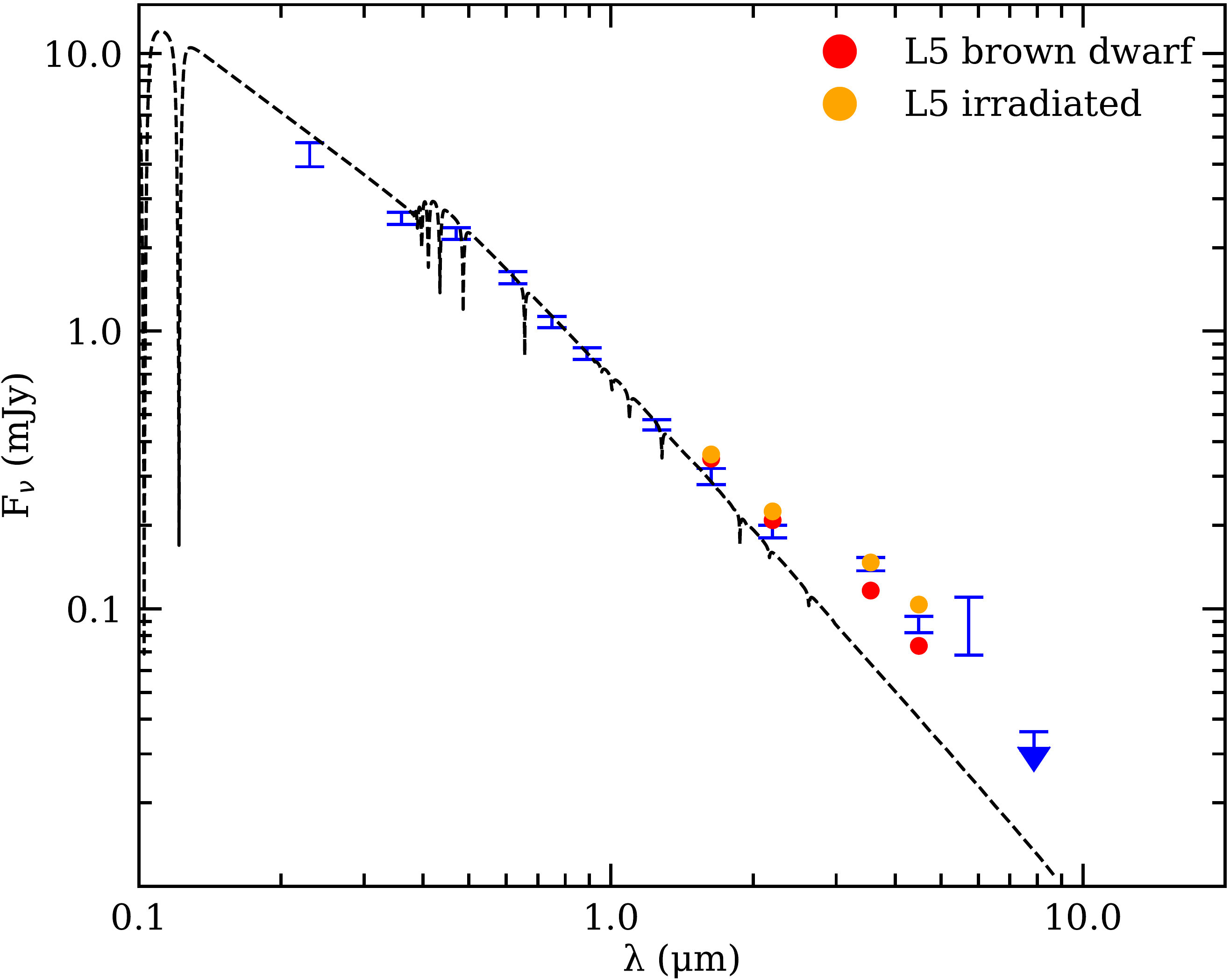}
\vskip 0mm
\caption{The photometric spectral energy distribution of PG\,0010$+281$.  Blue error bars show {\em GALEX} near-ultraviolet, optical SDSS $ugriz$, and {\em Spitzer} IRAC fluxes (measured here independently).  The stellar photosphere is plotted as a black dashed line, and modelled using a 27\,000\,K pure hydrogen atmosphere white dwarf \citep{2010Koester}.  Also plotted are models that include the addition of flux from an L5-type brown dwarf, both without (red circles) and with (orange circles) the influence of stellar irradiation.  The brown dwarf fluxes are derived from an extensive study of M, L, and T dwarfs using {\em Spitzer} IRAC \citep{2006Patten}, and the effects of irradiation were estimated based on the observed IRAC variability for WD\,0137--349 \citep{2015Casewell}.  The IRAC 3.6$\upmu$m and 4.5$\upmu$m data for PG\,0010+281 were 10--15\,min apart, where an irradiated companion could change aspect somewhat for relatively short orbital periods.}
\label{fig:pg0010_irsed}
\end{figure}

A more widely separated, self-luminous companion is possible, as it would not lead to photometric or radial velocity variability.  The existing $K$-band image of PG\,0010$+281$ \citep{2015Xu} should rule out an equally luminous companion at a projected separation beyond roughly 33\,AU, which is $1/3$ of the 0.8\,arcsec FWHM reported in the observation, at a distance of 134\,pc based on the \textit{Gaia} DR3 parallax of $7.47\pm0.06$\,mas.  Figure~\ref{fig:pg0010_irsed} shows the spectral energy distribution of PG\,0010+281 from the ultraviolet through the infrared, including {\em Spitzer} IRAC photometry performed and analysed here independently.  The data at 5.7\,$\upmu$m are ignored as the errors are too large to be reliable, and furthermore the IRAC image appears possibly extended, and thus may contain light from a background galaxy.  Models for an irradiated and non-irradiated L5 dwarf are broadly consistent with the observed photometry, given an uncertain companion temperature and radius.  Keeping in mind that the two, shorter-wavelength IRAC observations are non-simultaneous, and the fact that the effects of irradiation are likely to be diverse across individual systems \citep{2022Lee}, this model seems plausible solely on the basis of the infrared photometric excess.  

However, as discussed in the previous sections, the constraints on radial velocity, photometric modulation, and astrometry only allow a narrow and finely-tuned range of orbital configurations.  A substellar companion, such as those known to orbit white dwarfs, would have to be in a nearly face-on (and short-period) orbit to eschew detection in all three of these observable metrics, and is therefore unlikely.

%%%FIGURE 8%%%

\subsubsection{A debris disk?}

While substellar companions cannot be completely ruled out, if the relatively low white dwarf mass is set aside, then the results favour a debris disk as the cause of the infrared excess.  In this case, the thermal emission signature is subtle compared to many white dwarfs that reprocess up to 3 per cent of the incident stellar radiation \citep{2015Rocchetto}, but such faint disks are well represented in the population of metal-polluted white dwarfs observed with {\em Spitzer} \citep{2010Farihi, 2014Bergfors}.

It is now well understood that circumstellar material orbiting white dwarfs is not always contained within the Roche radius, nor always or only in an optically thick configuration \citep{2018Farihi_b, 2020Swan}.  At the time of discovery, \citet{2015Xu} interpreted the infrared excess towards PG\,0010+281 as a viable dust disk within the Roche limit using the older, canonical model of a flat and opaque configuration, as in the rings of Saturn.  That study appealed to the shielding of dust within a vertically flat structure, to argue that 1200\,K grains can be present near 0.5\,R$_\odot$, and concluding such a disk was possible.  However, fully-exposed blackbody grains would attain $T>2200$\,K at 1\,R$_\odot$ from the white dwarf, and thus there can be no solids present to provide the hypothesized shielding \citep{2012Rafikov}.  An independent study has concluded that, in general, all material within the Roche limit would be rapidly sublimated by a 26\,500\,K white dwarf because no dust can arrive intact or otherwise be produced in this region \citep{2021Steckloff}.

For an infrared excess with emission temperatures in the range 1000--1300\,K, a white dwarf temperature 26\,500\,K, and a stellar radius 0.0146\,R$_\odot$, blackbody dust should be located at radii 3.0--5.1\,R$_\odot$, and thus orbit comfortably distant from the Roche limit.  This debris disk model can be interpreted in a rather straightforward manner, where all material closer to the star is gaseous and thus unseen in the infrared.  Poynting-Robertson drag on the inner portion of the disk near 3\,R$_\odot$ would remove typical $1-10$\,$\upmu$m grains on roughly $4-40$\,yr timescales and a pileup should occur until fully depleted.  If confirmed, such a debris disk would be yet another example of dust orbiting well outside the Roche limit \citep{2017Farihi, 2020Vanderbosch, 2021Guidry, 2022Farihi}, and thus another example that the canonical model of flat and opaque rings \citep{2003Jura} is either incorrect or not the whole picture for those stars with prominent infrared excess in {\em Spitzer} and {\em WISE} \citep{2018Farihi_b, 2020Swan}.  And while heavy elements are not detected in optical spectra of PG\,0010+281, it is not surprising given the relatively high temperature and corresponding opacity in the white dwarf atmosphere.  A prediction of the debris disk model supported here, is that metal pollution should be apparent in ultraviolet observations, which are more sensitive for stars of similar temperature \citep{2014Koester}.

Observations with {\em JWST} MIRI can in principle confirm the dust with the detection of solid-state features such as those typical of planetary silicates \citep{2009Jura}.  However, if there is both dust and a (wide or otherwise hidden) substellar companion, as in SDSS\,J155720.77+091624.6, then the interpretation of the infrared spectra may be complicated.

\subsubsection{A cannibalized companion?}

There is tension between the low white dwarf mass and the old total age it implies, versus the expectations of single star evolution, where known white dwarfs hosting debris disks are essentially a single star (or distant binary) population \citep{2019Wilson}.  And although these are not mutually exclusive possibilities in terms of the circumstellar architecture, the disparate age indicators cannot be reconciled easily.  One possible way to resolve this tension, is if the white dwarf destroyed one or more giant planets on the AGB, which enhanced mass loss during their final death spiral (i.e.\ a fatal common envelope (\citealt{2012Mustill, 2013Nordhaus, 2013Soker}).  

The cooling age for a 0.57\,M$_{\odot}$ white dwarf with $T_{\rm eff}=27\,000$\,K is negligible ($<15$\,Myr).  Under an assumption of single star evolution for such a low remnant mass, however, there are essentially no reliable empirical constraints on the zero age main-sequence progenitor.  As mentioned above, IFMRs for a 0.57\,M$_{\odot}$ white dwarf predict main-sequence progenitors with hydrogen-burning lifetimes approaching or exceeding a Hubble time, thus suggesting the stellar evolution of PG\,0010+281 was not typical of isolated stars.

While somewhat speculative, it would be beneficial to simulate the effects of planetary engulfment on post-main sequence evolution, especially in terms of mass loss.  It is well known that main-sequence stars commonly host multiple planets that orbit sufficiently close to be engulfed during the post-main sequence.  The basic properties of white dwarfs, such as atmospheric composition, structure and mass of the outer layers, total remnant mass, and magnetism might all be dependent on such processes.  Thus, the effect of planets on stellar evolution may be a key issue in white dwarf formation and subsequent evolution.

\section{L dwarf survivors of giant encounters}

To date, the only substellar survivors of one or both phases of giant stellar evolution appear to be brown dwarfs with masses in the range of roughly $50-70$\,M$_{\rm Jup}$, and with spectral types consistent with L dwarfs (nine currently known, including GD\,1400B; \citealt{2022Zorotovic}).  In contrast, at wider orbital separations, there is a small number of substellar companions to white dwarfs that are unambiguously cooler or less massive, in addition to widely bound L-type dwarfs.  A possibly incomplete list of wide substellar companions to white dwarfs, where a common envelope would not have developed, includes: the prototype L dwarf GD\,165B \citep{1988Becklin}, the L dwarf PHL\,5308B \citep{2009Steele}, the T dwarfs LSPM\,J1459+0851B \citep{2011Dayjones} and PSO\,J058.9855+45.4184B \citep{2020Zhang}, the likely Y-type dwarf WD\,0806-661B \citep{2012Luhman,2014Luhman}, and an old, metal-poor T dwarf Wolf\,1130C, which is bound to a post-common envelope (11.9\,h) binary consisting of an M subdwarf and an unseen, massive white dwarf Wolf\,1130AB \citep{2018Mace}.

One possible exception to this is the Jupiter-sized body in a 1.4\,d orbit around WD\,1856+534 \citep{2020Vanderburg}, assuming that it did actually emerge from a common envelope.  That scenario is uncertain for a few reasons, one of which is the fact that post-common envelope, low-mass stellar or substellar companions at these longer orbital periods are rare or non-existent \citep{2011Nebot}.  Another potential complication is that there are not one, not two, but three stars in this system \citep{1999Mccook,2020Vanderburg}, and more than one study has suggested this object arrived recently by Kozai-Lidov migration \citep{2021Munoz,2021Oconnor,2021Stephan}.  Other solutions that invoke the common envelope require fine tuning \citep{2021Lagos} and are thus far less likely.

Therefore, at face value, the evidence suggests that a mass of around 50\,M$_{\rm Jup}$ is a robust benchmark for survival of the common envelope.  This strongly implies that all objects of planetary masses as represented by the solar system, are destroyed during the RGB or AGB if engulfed, and merge with the central star.  These merging events, of objects up to a few tens of M$_{\rm Jup}$, may have observational implications for the final mass and properties of the white dwarf remnant, including the potential emergence of a magnetic field.

\section{Summary and Conclusions}

New data are presented and analysed for GD\,1400 and PG\,0010$+281$, both white dwarfs where closely-orbiting substellar companions were previously suspected.  The radial velocity and photometric search for a substellar companion to PG\,0010+281 did not identify any candidates, and relatively good constraints are established for short periods that are typical of post-common envelope binaries.  The infrared spectroscopy presented here does not conclusively distinguish between a cool companion and dust emission, but photospheric Br$\upgamma$ is detected which likely favours the latter.  As a result, it appears the infrared excess measured for PG\,0010$+281$ is more consistent with a single star and circumstellar dust, but its mass and age appear potentially inconsistent with isolated stellar evolution.

In contrast, GD\,1400A and B are successfully characterised in this study, including orbital parameters, substellar companion mass, and its $R\approx9000$ $K$-band spectrum.  Derived from comparisons with synthetic templates, the benchmark brown dwarf appears relatively warm at $T_{\rm eff}\approx2100$\,K, and also rather young with an implied age of less than around 1\,Gyr according to evolutionary models for the calculated mass of 68\,M$_{\rm Jup}$.  The study notes a dearth of T-type and later brown dwarfs as companions to white dwarfs in post-common envelope systems, whereas cooler and less massive brown dwarfs clearly form and survive at wide separations.  This suggests a crude benchmark of 50\,M$_{\rm Jup}$ for a companion to survive engulfment on the RGB (or both RGB and AGB) common envelope evolution. 

\section*{Acknowledgements}
Dedicated to the memory of Prof. T. R. Marsh (1961-2022).

The authors thank O.~Toloza for assistance with white dwarf atmospheric models, and the referee for useful comments that improved the manuscript.  This work is based on author PI and archival data, including software support, from several facilities (all programme IDs are provided in the manuscript): the European Southern Observatory and Science Archive Facility, the AURA Gemini North Observatory via remote access to the Canada-France-Hawaii Telescope, the Isaac Newton and William Herschel Telescopes, the Multiple Mirror Telescope, the Keck Observatory Archive, the NASA Infrared Telescope Facility, and the NASA/IPAC Infrared Science Archive.  This work has made use of data from the ESA {\it Gaia} satellite, processed by DPAC, and the {\em TESS} mission, whose funding is provided by the NASA Explorer Program.  This research made use of Astropy,\footnote{http://www.astropy.org} a community-developed core Python package for Astronomy \citep{2013Astropy, 2018Astropy}.  NW was supported by a UK STFC studentship hosted by the UCL Centre for Doctoral Training in Data Intensive Science.  JF acknowledges STFC grant ST/R000476/1 and similarly TRM was supported by grant ST/T000406/1.  JJH acknowledges support through TESS Guest Investigator Programs 80NSSC20K0592 and 80NSSC22K0737.

\section*{Data Availability}

The data underlying this article will be shared on reasonable request to the corresponding author.

%\begin{thebibliography}{}

\bibliographystyle{mnras}
\bibliography{references}

\bsp    % typesetting comment
\label{lastpage}
\end{document}